\documentclass[twocolumn,secnumarabic,amssymb, nobibnotes, aps, pra,superscriptaddress]{revtex4-2}
\usepackage[colorlinks=true,linkcolor=blue,citecolor=blue,urlcolor=blue]{hyperref}
\usepackage[separate-uncertainty=true]{siunitx}
\usepackage{graphicx}
\usepackage{gensymb}
\usepackage{amsmath,bm}
\usepackage{braket}
\usepackage[normalem]{ulem}
\usepackage[shortlabels]{enumitem}
\makeatletter
\let\cat@comma@active\@empty
\makeatother

\newcommand{\micro}{\si{\micro}}

\begin{document}

\title{Source noise suppression in attosecond transient absorption spectroscopy by edge-pixel referencing}

\author{Romain G\'{e}neaux}
\email{romain.geneaux@cea.fr}
\affiliation{Department of Chemistry, University of California, Berkeley, CA 94720, USA.}
\affiliation{Université Paris-Saclay, CEA, CNRS, LIDYL, 91191 Gif-sur-Yvette, France.}

\author{Hung-Tzu Chang}
\affiliation{Department of Chemistry, University of California, Berkeley, CA 94720, USA.}

\author{Adam M. Schwartzberg}
\affiliation{Molecular Foundry, Lawrence Berkeley National Laboratory, Berkeley, CA 94720, USA.}

\author{Hugo J. B. Marroux}
\email{hugo.marroux@epfl.ch}
\affiliation{Laboratoire de Spectroscopie Ultrarapide (LSU) and Lausanne Centre for Ultrafast Science (LACUS), \'{E}cole  Polytechnique  F\'{e}d\'{e}rale de  Lausanne, ISIC, FSB, Station 6, CH-1015 Lausanne, Switzerland.}

\begin{abstract}
Attosecond transient absorption spectroscopy (ATAS) is used to observe photoexcited dynamics with outstanding time resolution. The main experimental challenge of this technique is that high-harmonic generation sources show significant instabilities, resulting in sub-par sensitivity when compared to other techniques. This paper proposes edge-pixel referencing as a means to suppress this noise. Two approaches are introduced: the first is deterministic and uses a correlation analysis, while the second relies on singular value decomposition. Each methods is demonstrated and quantified on a noisy measurement taken on $\text{WS}_2$ and results in a fivefold increase in sensitivity. The combination of the two methods ensures the fidelity of the procedure and can be implemented on live data collection but also on existing datasets. The results show that edge-referencing methods bring the sensitivity of ATAS near the detector noise floor. An implementation of the post-processing code is provided to the reader.
\end{abstract}

\maketitle

\section{Introduction}
Transient absorption experiments are now frequently performed in the extreme ultraviolet (XUV) and X-ray regimes, taking advantage of the element specificity and the sensitivity to local structural and electronic environments offered by these radiations. The earliest of these experiments emerged more than 30 years ago \cite{Epstein1983,Murakami1986} and used laser-produced plasmas as X-ray sources. Now, experiments are commonly performed on gas, liquid and solid targets, for instance at large instrument facilities such as synchrotrons \cite{Cavalleri2005,Bressler2009} and free-electron lasers \cite{Obara2017}. While these sources offer freely tunable and high flux X-ray radiation, they suffer from a probe bandwidth which is intrinsically narrow (< 1 eV) compared to absorption edges. This forces experimentalists to scan the central frequency of the probe pulse at a given time delay, resulting in prohibitively long acquisition times. On the other hand, high-order harmonic generation (HHG) sources provide lower flux but extremely wide bandwidth radiation routinely covering > 30 eV. This allows to perform attosecond transient absorption spectroscopy (ATAS), in which multiple absorption edges are covered in one laser shot and where time resolutions are on the order of attoseconds \cite{Geneaux2019,Gallmann2013}. 

While ATAS is becoming an increasingly useful technique to study both molecular \cite{Saito2019,Marroux2020} and solid-state processes \cite{Buades2018,Volkov2019a}, the sensitivity of hitherto published experiments has been limited. The detection limit of these experiments expressed in a change of optical density is typically larger than $\Delta\text{OD} = 10^{-3}$, which has to be compared to experiments performed in the visible and mid-infrared spectral ranges, where optical density changes of $\Delta\text{OD} = 10^{-5} - 10^{-6}$  can be observed \cite{Lang2018,Oppermann2019}. This limitation is mainly due to the high non-linearity of the HHG process, which results in strongly correlated spectral noise. 

Recently, two experimental approaches were proposed to tackle this issue. Volkov et al.~proposed to use the strong correlation between the fluctuations in the intensity of the driving laser and shifts in the XUV spectrum to correct for correlated noise, which improved their detection limit by a factor of two \cite{Volkov2019}. Two other works implemented a parallel synchronized measurement of a reference spectrum, which allows for normalization of the probe spectrum. This was performed using either a single spectrometer measuring both transmitted and reference XUV beams \cite{Stooss2019}, or in a dual spectrometer geometry \cite{Willems2020}. In the latter experiment, sensitivities below 10 mOD were obtained, showing the promises of the approach. However, because this requires additional hardware and reduces the available XUV flux, it is not readily applicable to any experiment.

Here we take a different approach and explore two different post-processing procedures based on edge-referencing \cite{Feng2017,Robben2020}. The methods use a general noise suppression scheme taking advantage of strong correlations in the probe spectra, which are first characterized on a calibration dataset. We present two complementary methods and demonstrate their efficiency in an attosecond transient reflectivity experiment, measured on a sample of $\text{WS}_2$. The edge-referenced data obtained using either methods shows a drastic fivefold increase in sensitivity which in turn allows for the observation of features previously hidden by spectral fluctuations. Our approach is significant in four respects: (1) it radically filters out the correlated part of the noise without requiring any new hardware, 
(2) it does not require sacrificing a part of the XUV flux for reference detection, and is therefore applicable to virtually any ATAS experiments, (3) the combination of the two methods ensures that no artefacts are introduced during the edge-referencing and (4) the noise calibration step can be constructed from information contained in existing datasets, which makes it valuable to analyze already measured data. We provide a code implementing the procedure which can be readily applied to existing datasets.

\section{Attosecond transient absorption and reflectivity spectroscopy}

\subsection{Formalism}
We consider the most frequent ATAS scheme in which the sample is excited by a pump field in the visible/near-infrared range (NIR), and probed by a delayed XUV pulse. It is customary to measure the spectrum of the transmitted light twice, with and without the pump beam. From the pump-on $I^{1}$ and pump-off $I^{0}$ measurements the change in optical density is obtained as:
\begin{equation}
\begin{aligned}
\Delta \text{OD} &=-\text{log}_{10}\frac{I^{1}}{I^{0}}\approx-\text{log}_{10}\frac{I^1_\text{XUV}+\chi^{(3)} I^1_\text{XUV} I_\text{NIR}}{I^0_\text{XUV}}\\
&= -\text{log}_{10}\frac{I^{1}_\text{XUV}}{I^{0}_\text{XUV}} + \Delta \text{OD}_\text{signal}\cdot F,
\label{dOD}
\end{aligned}
\end{equation}
where the measured $\Delta\text{OD}$ was factorized into additive noise, multiplicative noise $F$, and the signal of interest $\Delta\text{OD}_\text{signal}$. Optical densities are preferred to transmission because $\Delta \text{OD}_\text{signal}$ scales linearly with concentration or thickness. 

Multiplicative noise is inherently much smaller than the signal itself and will henceforth be neglected \cite{Robben2020}. Additive noise disappears provided that $I^1_\text{XUV}=I^0_\text{XUV}$, meaning that the XUV fluctuations between the pump-on/off measurements must be negligible and the measurement infinitely precise. It is therefore customary to measure pump-on and pump-off spectra at the highest frequency possible in order to limit the effect of additive noise. Yet, additive noise is never fully mitigated because the XUV probe has limited stability:  it is generated via HHG, whose non-linearity amplifies the driving laser intensity, pointing and mode fluctuations \cite{Erny2011, Volkov2019}. For short driving pulses as used here, carrier-envelope phase noise is an additional source of noise. In the following, two procedures are proposed to suppress this XUV noise so that the measurement becomes limited by the readout noise of the detector, similarly to visible and mid-infrared transient absorption. 

\begin{figure*}
 \includegraphics[width=\linewidth]{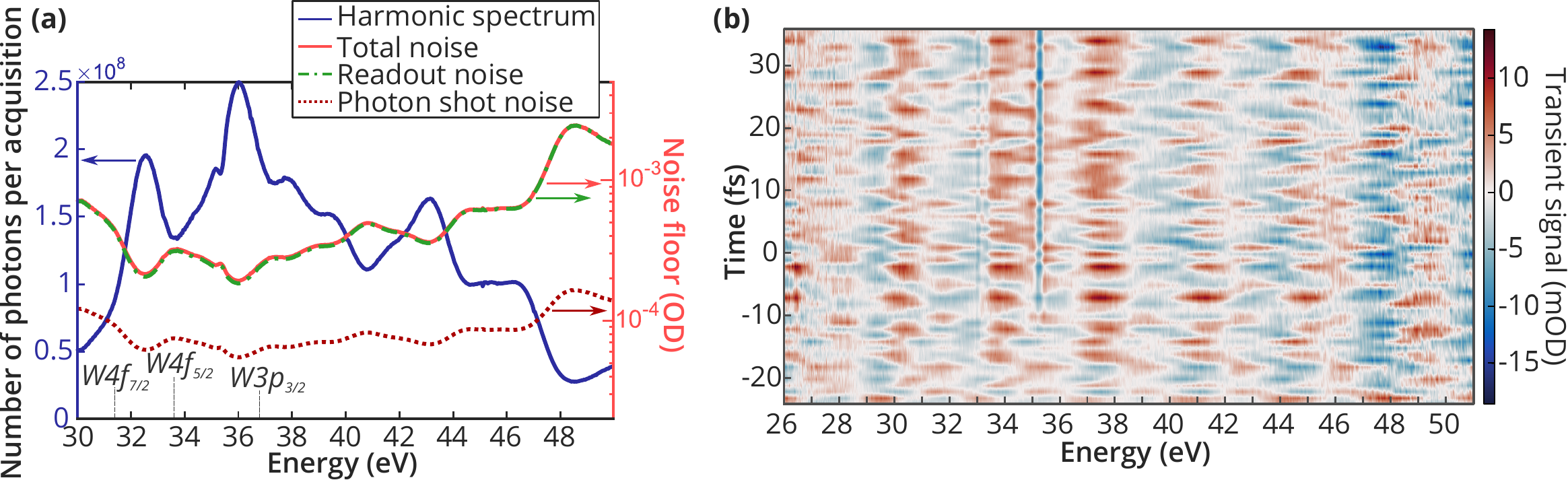}
 \caption{\label{fig1} \textbf{(a)} Noise floor of the experiment: The XUV spectrum (blue) is shown in terms of number of photon per acquisition with 500 ms exposure. The noise floor in units of optical density is shown on the right axis for the photon shot noise (dotted red line), camera readout noise (dash-dotted green line) and total noise (full red line). Relevant tungsten core-levels are indicated at the bottom of the figure. \textbf{(b)} Raw transient reflectivity spectrogram, defined with Equ.~\ref{dOD} and measured with steps of 1~fs.}
\end{figure*}
\subsection{Experiment}

To demonstrate our method, we use a dataset of attosecond pump-probe measurement obtained in bulk $\text{WS}_2$. Carrier-envelope phase stabilized few-cycle (5~fs) pulses of visible/near-infrared light are used to photoexcite the sample and to generate the broadband XUV radiation which serves as probe. Using krypton as a generating medium, the XUV spectrum is optimized to overlap with the 4f (31.4/33.6 eV) and 5p (36.8 eV) core-levels of tungsten, as shown on Fig.~\ref{fig1}(a). The XUV pulse probes the transitions from these core-levels to the valence and conduction band regions. The sample was produced by plasma-enhanced atomic layer deposition of 40 nm thick tungsten oxide on a silicon wafer capped with 250 nm thick SiO$_2$, and subsequently converting the tungsten oxide to WS$_2$ by reacting the deposited tungsten oxide with H$_2$S and Ar as buffer gas at 550 $^{\circ}$C \cite{Kastl2017}. As the sample is opaque to the XUV, a reflectivity \cite{Kaplan2018} rather than an absorption experiment was performed. While dynamics are in this case usually reported in terms of reflectivity changes, we will keep using changes of optical densities, following Equation \ref{dOD}. Both quantities are proportional to each other in the small signal limit, but optical densities are a more common unit to report experimental sensitivities - we therefore keep using it for the sake of generality.\\

The XUV probe reflected from the sample is dispersed and measured by a 1340$\times$400 pixels CCD detector (PIXIS 400B, Princeton Instruments) cooled to -40{ }\degree{C}. We begin by determining the noise floor of the experiment, which comprises the detector dark noise (due to thermally generated electrons in the CCD), readout noise (arising when measuring the voltage induced by the electronic charge) and the photon shot noise (associated with the random arrival of photons on the sensor). With the camera settings used (500 ms exposure, 2MHz readout speed), the average signal is $1.5 \times 10^5$ counts/channel, the readout noise is measured to be $79$ counts/channel RMS and the dark noise is $<0.1$ counts/pixel, which is negligible. The photon shot noise is $\sqrt{N_\text{photon}}$ with $N_\text{photon}$ the number of detected photons converted from measured counts using the camera quantum efficiency, gain and number of electrons generated as a function of photon energy. These noises are propagated using the optical density equation and shown in Fig. \ref{fig1}(a). We see that readout noise is the major contribution of the noise floor which lies between 200 and 500~$\micro\text{OD}$ in the tungsten core-level region. 

Fig.~\ref{fig1}(b)~shows the raw pump-probe scan where each time delay was averaged 40 times. The only clear feature in the raw data is a narrow negative feature at 35.39 eV as well as two weaker signals at slightly lower photon energies. The atomic-like narrow linewidth of the signal and the fact that it starts appearing at -10 fs suggests a core-excitonic nature \cite{Moulet2017,Geneaux2020,lucchini2020unravelling}. Indeed, at negative delays, the role of the pump and probe pulses are reversed: signals in these regions are due to an XUV-triggered dynamics probed by the NIR field. At positive delays on the other hand, we expect signals triggered by the NIR pulse, which has an energy of $\hbar\omega_\text{pump}=1.2-2.5$ eV. This is enough to photoinject carriers across the $\sim$1.35 eV bandgap \cite{Kuc2011} of $\text{WS}_2$. Such carriers would give signals in the valence and conduction band regions \cite{Kaplan2018,Zurch2017} which here lie in the 36-41 eV range. The experimental noise of this dataset is too large to distinguish these features. In addition, the transient spectrum displays a very structured noise, which is typical of data taken with poor HHG stability. In the following section we will use edge-pixel referencing methods to retrieve the hidden information. 

\section{Methods}
\subsection{Edge-pixel referencing}
The XUV spectrum used here probes changes across 25 eV at once. This large bandwidth is a unique asset of HHG-based sources, but is actually much larger than the region where pump-induced signals are expected. Even considering excitonic effects and non-linear excitation, the pump pulse will typically modify the XUV spectrum at most a few eVs away from the energies of the core-levels used in the probe step. Hence, a large part of the XUV spectrum does not contain relevant information for the chemical or physical dynamics investigated. 

As the spectral noise of the probe pulse is strongly correlated, the intensity fluctuations of the signal-free region contains information on the fluctuations in the spectral region where the signal is located.  This means that the regions without signal - hereinafter referred to as \textit{edge-pixels} regions -  can be used to remove fluctuations in the regions with signal - now called the \textit{signal-pixels} region. This idea was recently used in IR spectroscopy by Robben et al.~\cite{Robben2020} who adapted the procedure originally developed by Feng et al.~\cite{Feng2017} for dual detector referencing. 
\begin{figure*}
 \includegraphics[width=\linewidth]{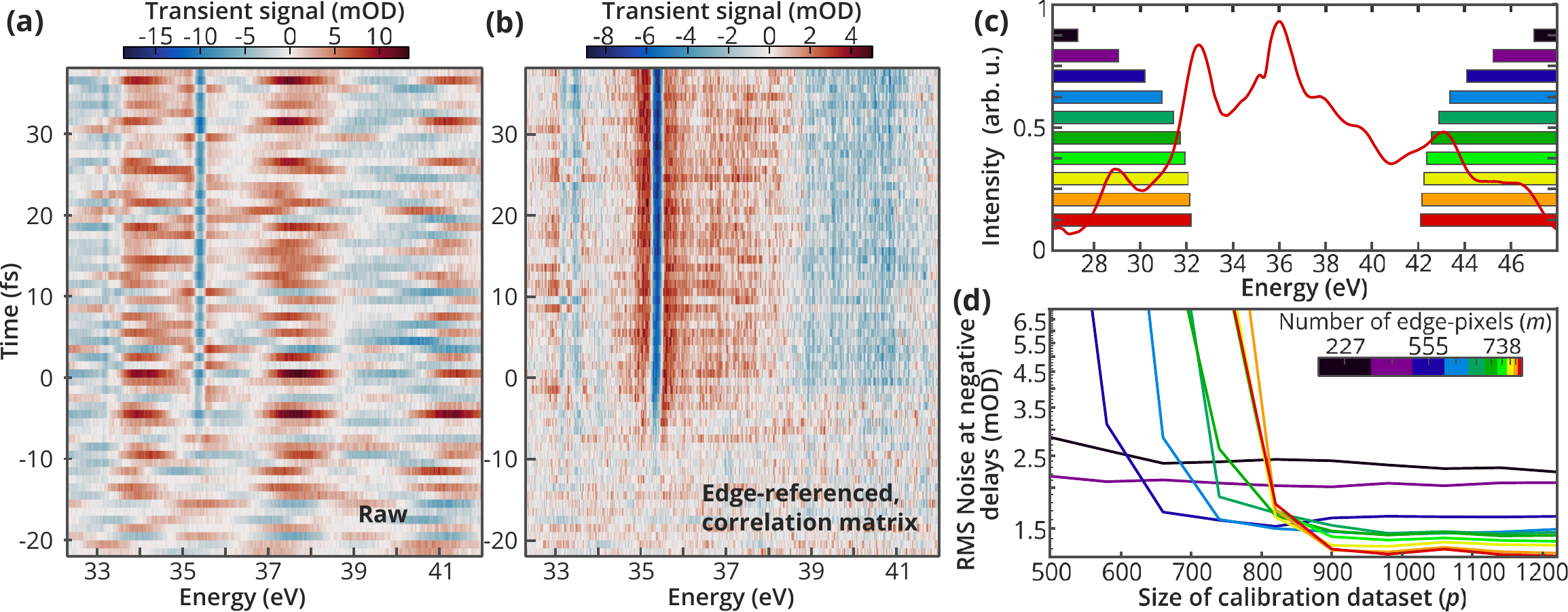}
 \caption{\label{fig2} \textbf{(a)} Raw transient spectrum in the signal region. \textbf{(b)} Edge-referenced transient spectra obtained using the correlation matrix method and edge-pixels in the regions 26-32 and 42-51~eV. \textbf{(c)} Average reflected spectrum (red line) together with edge-pixel regions (blue shaded areas). Darker and lighter regions correspond to fewer and more edge-pixels, respectively. \textbf{(d)} Root mean square of the noise at negative delays for the different edge-pixel regions shown in (c) as a function of the size of the calibration data set $p$.}
\end{figure*}
\subsection{First approach: correlation matrix}
In this approach, a correlation matrix is used to map the XUV fluctuations in the edge-pixels onto the signal-pixels. We note $n$ the number of pixels in the spectrum, and $m$ the number of pixels chosen as edge-pixels, with $m<n$. If $N_t$ is the number of pump-probe delays, $N_t$ pairs of pump-on and pump-off spectra are obtained in the experiment, yielding the raw transient absorption $\Delta \text{OD}_\text{measured}$ which is a $(N_t\times n)$ matrix. $\Delta \text{OD}^\text{edge}_\text{measured}$ is a $(N_t\times m)$ matrix - a subset of $\Delta \text{OD}_\text{measured}$ - used to correct the noise in the entire measurement region:
\begin{equation}\label{correction}\Delta OD_\text{referenced} = \Delta OD_\text{measured} - \Delta OD^\text{edge}_\text{measured}\cdot\mathbf{B}
\end{equation}
$\mathbf{B}$ is the $(m \times n)$ correlation matrix that is measurement specific and needs to be constructed for each experiment. The first step is to compute $\mathbf{B}$. To this end, we need a calibration dataset which is a series of probe spectra acquired while blocking the pump beam. Pairing these spectra two by two yields $p$ "blank" transient spectra, resulting in the $(p \times n)$ calibration matrix $\Delta \text{OD}_\text{calib}$. We follow Refs.~\cite{GeOE2013,Robben2020} to obtain the $\mathbf{B}$ matrix which minimizes the residual noise after referencing. It can be written in the following matrix form:
\begin{equation}\label{B} \mathbf{B} = \left({\Delta OD^\text{edge}_\text{calib}}^T\cdot    {\Delta OD^\text{edge}_\text{calib}}\right)^{-1}\cdot
\left({\Delta OD^\text{edge}_\text{calib}}^T\cdot    {\Delta OD_\text{calib}}\right), \end{equation}
where $^{-1}$ denotes the matrix inversion operation. Since $\Delta OD^\text{edge}_\text{calib}$ is a real, $(p\times m)$ matrix, we have $\text{rank}({\Delta OD^\text{edge}_\text{calib}}^T\cdot\Delta OD^\text{edge}_\text{calib}) = \text{rank}(\Delta OD^\text{edge}_\text{calib}) \leq \text{min}(p,m)$. Therefore, the inversion can be performed only if $p>m$. In other terms, the calibration dataset needs more observations as there are edge-pixels for Equ.~\ref{B} to be used. If this condition is met, computing $\mathbf{B}$ is extremely efficient. Finally we note that  Equ.~\ref{B} can also be formulated using covariances (as done in Ref.~\cite{GeOE2013}) using the fact that both $\Delta OD_\text{calib}$ and $\Delta OD^\text{edge}_\text{calib}$ have an expectation value of zero.

For experiments with long acquisition times and possibly varying noise structures, this calibration dataset can be updated periodically during the measurement. The transient data of Fig.~\ref{fig1} was taken without such calibration step and as such this approach is unavailable. However, the data was collected by alternating between pump-off and pump-on measurements. We can therefore use all the pump-off spectra as our calibration dataset for building the $\mathbf{B}$ matrix. This approach will not be as accurate as an experiment where the $\mathbf{B}$ matrix is periodically computed, but it has the advantage of being applicable to any dataset at the post-treatment stage. In our case, pairing the pump-off spectra two by two yields a total of $p=1220$ measurements for the matrix $\Delta\text{OD}_\text{calib}$. The next step is to define the \textit{edge-} and \textit{signal-pixels} regions in the spectrum. These ranges are the only parameters that must be chosen in the procedure. 

Figure \ref{fig2}(a-b) shows the correction obtained using the 766 pixels corresponding to energies ranging from 26-32 and 42-51~eV as edge-pixels. The improvement on signal quality is drastic, with most of the source noise disappearing and revealing salient transient signals. The sharp features at 33.17 eV, 33.5 eV and 35.38 eV become prominent which allow to identify their lineshape and confirm that they appear slightly before the pump-probe overlap. Broader features are now also resolved above 36 eV, at energies where valence and conduction band signals are expected. 

While the observed features are intriguing and might bring insight into the dynamics of $\text{WS}_2$, which has never been studied using transient XUV spectroscopy, their analysis goes beyond the scope of this work and will be the subject of a future publication. Nonetheless, the improvement of the data quality demonstrates the strength of the edge-referencing method. Its performance is now assessed quantitatively as a function of both the size of the calibration dataset $p$ and the number of edge-pixels $n$. The remaining noise level is defined as the root-mean square of the optical densities measured at negative delays ($t<-10$ fs). Figure \ref{fig2}(c-d) illustrate the evolution of the noise when varying the number of edge-pixels (darker blue shaded areas indicate less edge-pixels) and restricting the number of calibration $\Delta$OD spectra in $\Delta \text{OD}_\text{calib}$. The trend is very similar to the one identified in Ref. \cite{Robben2020}: larger edge-pixels regions increase the performance of the method, but require more calibration measurements to reach the lowest noise. Robben et al.~suggested that the noise asymptotic limit was attained for $p \gtrsim 10 \times n$, meaning that more calibration measurements might further increase the sensitivity of our experiment. An implementation of the procedure, which can be employed to process most ATAS datasets, is available at \cite{github-edge-referencing}.

\begin{figure*}
 \includegraphics[width=\linewidth]{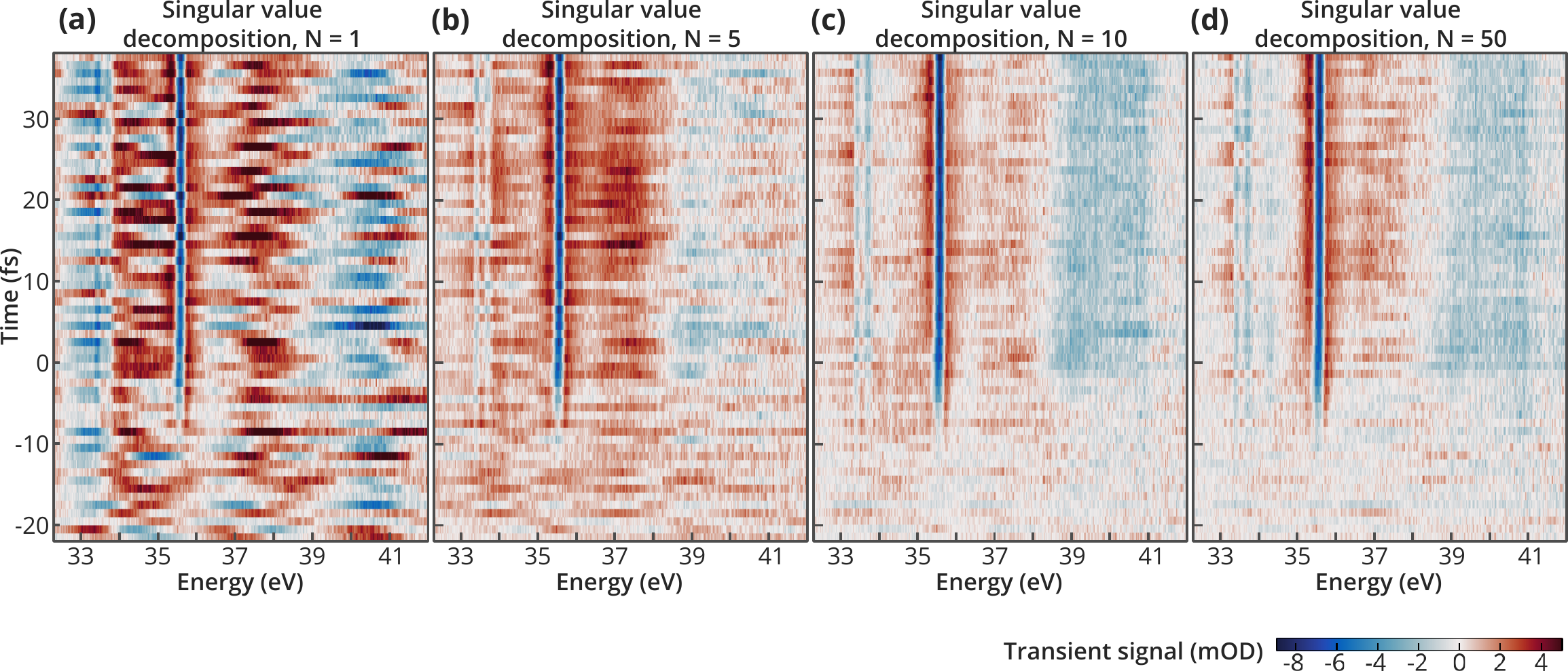}
 \caption{\label{fig3} Edge-referencing using SVD. Correction is applied with varying number of singular values. From left to right, N = 1, 5, 10 and 50. The chosen edge-pixels are the same as in Figure \ref{fig2} and correspond to 26-32 and 42-51~eV.}
\end{figure*}

\subsection{Second approach: singular value decomposition}
The correlation matrix method is optimal by definition: given the definition of the edge-pixels region and that enough spectra are collected  to build the $\mathbf{B}$ matrix, it provides the best noise correction \cite{Feng2017}. The only assumption is that no pump-induced signal is present in the edge-pixel region. However, should this assumption fail, it is important to stress that the correlation matrix approach can create artefacts. Going back to Equation \ref{correction}, we see that any unforeseen pump-induced signal appearing in $\Delta OD^\text{edge}_\text{measured}$ will be multiplied by $\mathbf{B}$ and directly transferred to other spectral regions. For XUV and X-ray spectroscopy the element edge structure restricts signals in defined edge and pre-edge regions but in the case of systems where multiple edges are covered, knowing where signals will appear is difficult, which could make it troublesome to rely on the $\mathbf{B}$ matrix correction alone.

For these reasons, we present a variant of the edge-referencing method which is less optimal to correct source noise but can be applied in a more controlled manner. Instead of computing $\mathbf{B}$, we perform a singular-value decomposition (SVD) of the calibration dataset. This yields the singular vectors, together with their respective singular values, which describe the structure of the noise. The number of significant singular values, i.e. values clearly higher than noise-related ones, helps to identify the dominant noise components. At the correction step, only the components corresponding to the $N$ highest singular values are fitted to the edge-pixels by performing a least-squares minimization. Results from the fit are then subtracted to $\Delta \text{OD}_\text{measured}$ in order to remove noise in the signal region. Thus, the impact of each singular vectors of the noise can be separated, and their shape can be compared with real features. The results are shown in Figure \ref{fig3} for $N$ varying from 1 to 50.

The spectra obtained using the SVD correction show a reduction of the structured noise as the number of singular values used increases. With $N>10$, similar features as the ones obtained using the correlation matrix methods become more pronounced, confirming that no artefact were induced by the $\mathbf{B}$ matrix correction. We note that for large $N$, the singular vectors become high frequency noise, not necessarily correlated between edge and signal regions. Regardless of this correlation, this high frequency noise will still be subtracted in the least-square fitting, resulting in a seemingly less noisy signal. It is important to remember this apparent noise reduction does not mean that the SVD correction performs better than the \textbf{B} matrix method, which is the optimal way of removing correlated source noise by definition. In the following section a quantitative analysis of the two correction schemes is discussed. 

\section{Discussion}
\begin{figure*}
 \includegraphics[width=\linewidth]{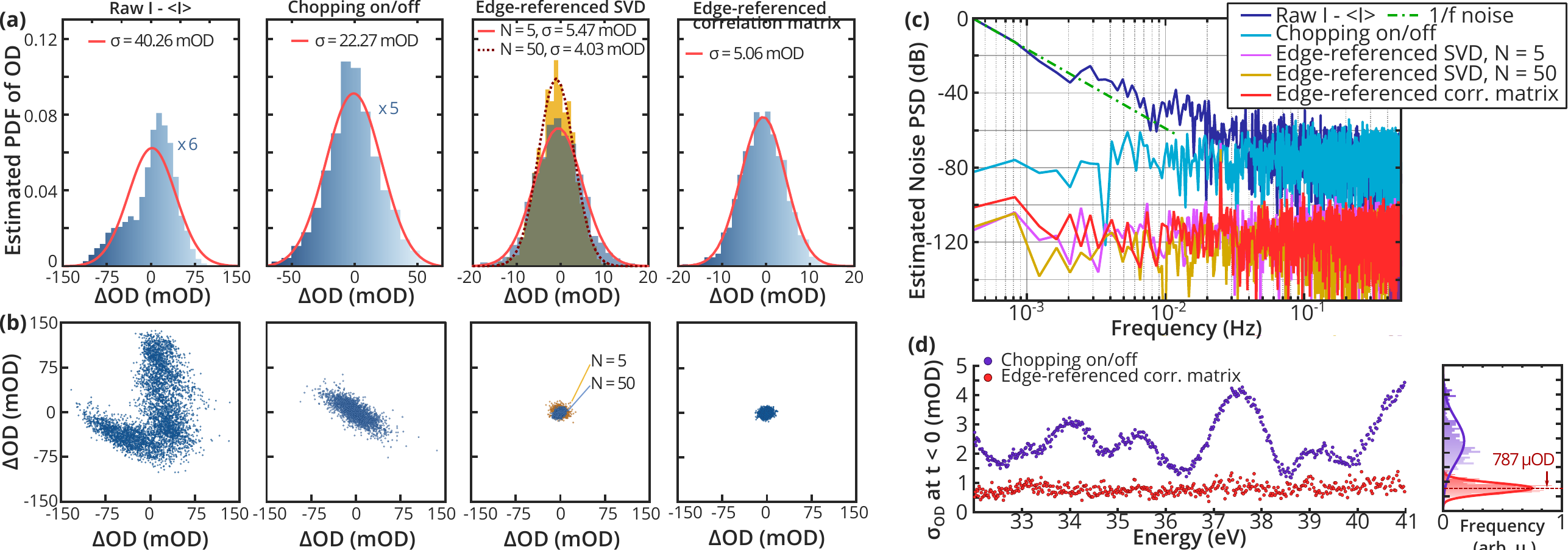}
 \caption{\label{fig4} \textbf{(a)} Probability density functions (PDF) of the single-shot $\Delta\text{OD}$ of one pixel for four experimental procedures: without chopping; with chopping, with SVD and with $\mathbf{B}$ matrix correction. Blue histograms are the measured noise distribution, while the pink curves are Gaussian fits of the histograms. Note that the x-axes have different scales, and that the first two PDFs were scaled to be displayed on the same y-axis. \textbf{(b)} Correlation between the measured $\Delta\text{OD}$ of two signal-free pixels. \textbf{(c)} Noise power density spectra. \textbf{(d)} Standard deviation of the transient optical density baseline at negative time delays with chopping (purple) and with $\mathbf{B}$ matrix correction (red). Corresponding histograms of the $\Delta\text{OD}$ are shown on the right. }
\end{figure*}

The performances of the two edge-referencing approaches are illustrated in Figure \ref{fig4}. In a signal-free region (below all W core-levels), we report the fluctuations of optical density for various analysis approaches. For each of them, we compute the probability density functions of the single-shot $\Delta\text{OD}$ at one pixel which allows us to compare the noise present in each analysis method.
The crudest measurement procedure used in the early days of ATAS consisted of measuring the time-dependent transmitted spectrum $I^{1}$ with the pump on at each time delay, and to divide each time step by the probe spectrum measured at a negative time delay (${<}I^{0}{>}$). Fig.~\ref{fig4}(a) shows that this approach results in a 40 $\Delta$mOD-wide PDF with a bi-nodal distribution. The noise power spectral density (Fig.~\ref{fig4}(c)) shows a very clear $\frac{1}f$ Flicker noise at low sampling frequencies. 
This noise is removed by modulating the pump using a mechanical shutter to alternate between pump-on and pump-off data at each time delay. This yields an approximately two-fold decrease in the width of the PDF and brings the distribution closer to a normal distribution. The power spectrum shows that noise contributions at lower frequencies than the chopping frequency (2 Hz) are removed. However, the correlation between two different signal free pixels (Fig.~\ref{fig4}(b)) shows that the remaining noise is highly correlated between different probe energies. 

When edge-referencing methods are applied, the PDFs further narrow to $\sigma<6$~mOD  and the noise spectrum is uniformly reduced by 40 dB. The correlation between pixels is removed which leads to an improvement of the baseline in the transient spectra. The final PDFs for both SVD and correlation matrix methods show a $\sim$4 to 5 fold width reduction compared to the sole chopping method. While the SVD method with $N=50$ gives the narrowest PDF, this is due to the filtering of uncorrelated high frequency singular values, as discussed above, and not to better performance compared to the correlation matrix approach. These plots represent sensitivities without averaging several scans. Since edge-referencing removed all correlated noise, the width of the PDF is expected to diminish as the square root of the number of averages.

Finally going back to the experimental transient spectrum of Figure \ref{fig1}, which was averaged 40 times, the performance of edge-referencing is ultimately quantified by studying the negative time delays (where no signal is present). As shown in Fig.~\ref{fig4}(d), the noise in the corrected spectra at negative delays is flat and has an average standard deviation of 787 $\micro\text{OD}$, which is consistent with a $\sqrt{40}$ improvement on the single shot PDF of Fig.~\ref{fig4}(a) and is remarkably close to the camera noise floor.

\section{Conclusion}
We have presented two general noise suppression schemes for ATAS that can be implemented either during data collection or applied at the post-treament stage. The two noise reduction schemes operate by correcting for the variations in source intensity in spectral regions where transient signal is located, using the fluctuations observed in the signal free regions. Using either a correlation matrix or a SVD method, noise originating from fluctuations in the spectrum of the XUV probe were reduced by a factor of five. This brings HHG-based ATAS closer to the detection limit of more established visible or mid-infrared transient absorption experiments. 

The correlation matrix scheme (code made available at  \cite{github-edge-referencing}) is particularly powerful, as it does not depend on any fitting and is therefore completely deterministic. The only two parameters that have to be decided by experimentalists are the spectral regions to be used for the referencing and the number of spectra to collect to accurately calibrate the $\mathbf{B}$ matrix. This is a decisive advantage compared to directly applying filters with arbitrary cutoff frequencies on the measured lineshapes.

While the number of spectra to be collected is easily decided by inspecting the improvement of the transient spectrum baseline, properly choosing the reference region can be the source of artefact if a region containing pump-induced signal is selected. In order to check that no artefact is introduced, a second method based on SVD can be used. If the reference region for the $\mathbf{B}$ matrix is properly selected the two methods show similar performances, although the $\mathbf{B}$ matrix method is significantly faster to implement.  

By implementing these methods in ATAS, the measurement quality is not limited by the instabilities of HHG anymore. With enough calibration datapoints, the detection limit will only be set by the detector. This might provide incentives to develop novel XUV detectors optimized for extremely low-readout noise. Not only does this allows us to observe dynamics with greater accuracy but it also relaxes the experimental constrains on flux and averaging times, which might prove key in the study of fragile solid state or even solvated molecules in liquid phase \cite{Kubin2018, Kleine2019}. These procedures can be employed in other regions of the XUV and X-ray spectrum and they will be particularly important for new generation sources operating at higher photon energies, whose lower photon fluxes makes accumulating statistics more difficult \cite{Popmintchev2012b,Johnson2018a,Barreau2020}. 

Finally, our approach can be directly applied to other types of broadband X-ray sources. Laser-plasma sources present similar noise characteristics \cite{Jonas2019}, albeit at much higher photon energies. Likewise, recent developments at free-electron laser sources reported spectral bandwidths up to 15~eV wide \cite{Duris2020,Driver2020}. Edge-referencing should therefore be directly applicable to such ultrashort and high-power X-ray sources and might become crucial in alleviating their inherently large intensity fluctuations.

\section*{Funding}
Investigations at UC Berkeley were supported by the W.M. Keck Foundation award No. 046300-002, the Army  Research  Office  Multidisciplinary University Research Initiative (MURI) grant No. WN911NF-14-1-03 and the U.S. Air Force Office of Scientific Research No. FA9550-19-1-0314. Work at the Molecular Foundry was supported by the Office of Science, Office of Basic Energy Sciences, of the U.S. Department of Energy under Contract No. DE-AC02-05CH11231.
H.J.B.M. received funding from the European Union’s Horizon 2020 research and innovation program under the Marie Skłodowska-Curie grant agreement No 801459 - FP-RESOMUS - and the Swiss National Science Foundation through the NCCR MUST. R.G. acknowledges funding from the LabEx PALM (ANR-10-LABX-0039-PALM).
\\
\section*{Acknowledgments}
We thank Dr. Lou Barreau for comments on the manuscript.

\noindent The authors declare no conflicts of interest.

\label{sec:refs}

\bibliography{library}

\begin{thebibliography}{36}%
\makeatletter
\providecommand \@ifxundefined [1]{%
 \@ifx{#1\undefined}
}%
\providecommand \@ifnum [1]{%
 \ifnum #1\expandafter \@firstoftwo
 \else \expandafter \@secondoftwo
 \fi
}%
\providecommand \@ifx [1]{%
 \ifx #1\expandafter \@firstoftwo
 \else \expandafter \@secondoftwo
 \fi
}%
\providecommand \natexlab [1]{#1}%
\providecommand \enquote  [1]{``#1''}%
\providecommand \bibnamefont  [1]{#1}%
\providecommand \bibfnamefont [1]{#1}%
\providecommand \citenamefont [1]{#1}%
\providecommand \href@noop [0]{\@secondoftwo}%
\providecommand \href [0]{\begingroup \@sanitize@url \@href}%
\providecommand \@href[1]{\@@startlink{#1}\@@href}%
\providecommand \@@href[1]{\endgroup#1\@@endlink}%
\providecommand \@sanitize@url [0]{\catcode `\\12\catcode `\$12\catcode
  `\&12\catcode `\#12\catcode `\^12\catcode `\_12\catcode `\%12\relax}%
\providecommand \@@startlink[1]{}%
\providecommand \@@endlink[0]{}%
\providecommand \url  [0]{\begingroup\@sanitize@url \@url }%
\providecommand \@url [1]{\endgroup\@href {#1}{\urlprefix }}%
\providecommand \urlprefix  [0]{URL }%
\providecommand \Eprint [0]{\href }%
\providecommand \doibase [0]{https://doi.org/}%
\providecommand \selectlanguage [0]{\@gobble}%
\providecommand \bibinfo  [0]{\@secondoftwo}%
\providecommand \bibfield  [0]{\@secondoftwo}%
\providecommand \translation [1]{[#1]}%
\providecommand \BibitemOpen [0]{}%
\providecommand \bibitemStop [0]{}%
\providecommand \bibitemNoStop [0]{.\EOS\space}%
\providecommand \EOS [0]{\spacefactor3000\relax}%
\providecommand \BibitemShut  [1]{\csname bibitem#1\endcsname}%
\let\auto@bib@innerbib\@empty
\bibitem [{\citenamefont {Epstein}\ \emph {et~al.}(1983)\citenamefont
  {Epstein}, \citenamefont {Schwerzel}, \citenamefont {Mallozzi},\ and\
  \citenamefont {Campbell}}]{Epstein1983}%
  \BibitemOpen
  \bibfield  {author} {\bibinfo {author} {\bibfnamefont {H.~M.}\ \bibnamefont
  {Epstein}}, \bibinfo {author} {\bibfnamefont {R.~E.}\ \bibnamefont
  {Schwerzel}}, \bibinfo {author} {\bibfnamefont {P.~J.}\ \bibnamefont
  {Mallozzi}},\ and\ \bibinfo {author} {\bibfnamefont {B.~E.}\ \bibnamefont
  {Campbell}},\ }\bibfield  {title} {\bibinfo {title} {{Flash-EXAFS for
  Structural Analysis of Transient Species: Rapidly Melting Aluminum}},\ }\href
  {https://doi.org/10.1021/ja00344a008} {\bibfield  {journal} {\bibinfo
  {journal} {Journal of the American Chemical Society}\ }\textbf {\bibinfo
  {volume} {105}},\ \bibinfo {pages} {1466} (\bibinfo {year}
  {1983})}\BibitemShut {NoStop}%
\bibitem [{\citenamefont {Murakami}\ \emph {et~al.}(1986)\citenamefont
  {Murakami}, \citenamefont {Gerritsen}, \citenamefont {{Van Brug}},
  \citenamefont {Bijkerk}, \citenamefont {Saris},\ and\ \citenamefont {{Van Der
  Wiel}}}]{Murakami1986}%
  \BibitemOpen
  \bibfield  {author} {\bibinfo {author} {\bibfnamefont {K.}~\bibnamefont
  {Murakami}}, \bibinfo {author} {\bibfnamefont {H.~C.}\ \bibnamefont
  {Gerritsen}}, \bibinfo {author} {\bibfnamefont {H.}~\bibnamefont {{Van
  Brug}}}, \bibinfo {author} {\bibfnamefont {F.}~\bibnamefont {Bijkerk}},
  \bibinfo {author} {\bibfnamefont {F.~W.}\ \bibnamefont {Saris}},\ and\
  \bibinfo {author} {\bibfnamefont {M.~J.}\ \bibnamefont {{Van Der Wiel}}},\
  }\bibfield  {title} {\bibinfo {title} {{Pulsed-laser irradiated silicon
  studied by time-resolved x-ray absorption (90-300 eV)}},\ }\href
  {https://doi.org/10.1103/PhysRevLett.56.655} {\bibfield  {journal} {\bibinfo
  {journal} {Physical Review Letters}\ }\textbf {\bibinfo {volume} {56}},\
  \bibinfo {pages} {655} (\bibinfo {year} {1986})}\BibitemShut {NoStop}%
\bibitem [{\citenamefont {Cavalleri}\ \emph {et~al.}(2005)\citenamefont
  {Cavalleri}, \citenamefont {Rini}, \citenamefont {Chong}, \citenamefont
  {Fourmaux}, \citenamefont {Glover}, \citenamefont {Heimann}, \citenamefont
  {Kieffer},\ and\ \citenamefont {Schoenlein}}]{Cavalleri2005}%
  \BibitemOpen
  \bibfield  {author} {\bibinfo {author} {\bibfnamefont {A.}~\bibnamefont
  {Cavalleri}}, \bibinfo {author} {\bibfnamefont {M.}~\bibnamefont {Rini}},
  \bibinfo {author} {\bibfnamefont {H.~H.}\ \bibnamefont {Chong}}, \bibinfo
  {author} {\bibfnamefont {S.}~\bibnamefont {Fourmaux}}, \bibinfo {author}
  {\bibfnamefont {T.~E.}\ \bibnamefont {Glover}}, \bibinfo {author}
  {\bibfnamefont {P.~A.}\ \bibnamefont {Heimann}}, \bibinfo {author}
  {\bibfnamefont {J.~C.}\ \bibnamefont {Kieffer}},\ and\ \bibinfo {author}
  {\bibfnamefont {R.~W.}\ \bibnamefont {Schoenlein}},\ }\bibfield  {title}
  {\bibinfo {title} {{Band-selective measurements of electron dynamics in
  VO$_{2}$ using femtosecond near-edge X-ray absorption}},\ }\bibfield
  {journal} {\bibinfo  {journal} {Physical Review Letters}\ }\textbf {\bibinfo
  {volume} {95}},\ \href {https://doi.org/10.1103/PhysRevLett.95.067405}
  {10.1103/PhysRevLett.95.067405} (\bibinfo {year} {2005})\BibitemShut
  {NoStop}%
\bibitem [{\citenamefont {Bressler}\ \emph {et~al.}(2009)\citenamefont
  {Bressler}, \citenamefont {Milne}, \citenamefont {Pham}, \citenamefont
  {ElNahhas}, \citenamefont {van~der Veen}, \citenamefont {Gawelda},
  \citenamefont {Johnson}, \citenamefont {Beaud}, \citenamefont {Grolimund},
  \citenamefont {Kaiser}, \citenamefont {Borca}, \citenamefont {Ingold},
  \citenamefont {Abela},\ and\ \citenamefont {Chergui}}]{Bressler2009}%
  \BibitemOpen
  \bibfield  {author} {\bibinfo {author} {\bibfnamefont {C.}~\bibnamefont
  {Bressler}}, \bibinfo {author} {\bibfnamefont {C.}~\bibnamefont {Milne}},
  \bibinfo {author} {\bibfnamefont {V.-T.}\ \bibnamefont {Pham}}, \bibinfo
  {author} {\bibfnamefont {A.}~\bibnamefont {ElNahhas}}, \bibinfo {author}
  {\bibfnamefont {R.~M.}\ \bibnamefont {van~der Veen}}, \bibinfo {author}
  {\bibfnamefont {W.}~\bibnamefont {Gawelda}}, \bibinfo {author} {\bibfnamefont
  {S.}~\bibnamefont {Johnson}}, \bibinfo {author} {\bibfnamefont
  {P.}~\bibnamefont {Beaud}}, \bibinfo {author} {\bibfnamefont
  {D.}~\bibnamefont {Grolimund}}, \bibinfo {author} {\bibfnamefont
  {M.}~\bibnamefont {Kaiser}}, \bibinfo {author} {\bibfnamefont {C.~N.}\
  \bibnamefont {Borca}}, \bibinfo {author} {\bibfnamefont {G.}~\bibnamefont
  {Ingold}}, \bibinfo {author} {\bibfnamefont {R.}~\bibnamefont {Abela}},\ and\
  \bibinfo {author} {\bibfnamefont {M.}~\bibnamefont {Chergui}},\ }\bibfield
  {title} {\bibinfo {title} {{Femtosecond XANES Study of the Light-Induced Spin
  Crossover Dynamics in an Iron(II) Complex}},\ }\href
  {https://doi.org/10.1126/science.1165733} {\bibfield  {journal} {\bibinfo
  {journal} {Science}\ }\textbf {\bibinfo {volume} {323}},\ \bibinfo {pages}
  {489} (\bibinfo {year} {2009})}\BibitemShut {NoStop}%
\bibitem [{\citenamefont {Obara}\ \emph {et~al.}(2017)\citenamefont {Obara},
  \citenamefont {Ito}, \citenamefont {Ito}, \citenamefont {Kurahashi},
  \citenamefont {Th{\"{u}}rmer}, \citenamefont {Tanaka}, \citenamefont
  {Katayama}, \citenamefont {Togashi}, \citenamefont {Owada}, \citenamefont
  {Yamamoto}, \citenamefont {Karashima}, \citenamefont {Nishitani},
  \citenamefont {Yabashi}, \citenamefont {Suzuki},\ and\ \citenamefont
  {Misawa}}]{Obara2017}%
  \BibitemOpen
  \bibfield  {author} {\bibinfo {author} {\bibfnamefont {Y.}~\bibnamefont
  {Obara}}, \bibinfo {author} {\bibfnamefont {H.}~\bibnamefont {Ito}}, \bibinfo
  {author} {\bibfnamefont {T.}~\bibnamefont {Ito}}, \bibinfo {author}
  {\bibfnamefont {N.}~\bibnamefont {Kurahashi}}, \bibinfo {author}
  {\bibfnamefont {S.}~\bibnamefont {Th{\"{u}}rmer}}, \bibinfo {author}
  {\bibfnamefont {H.}~\bibnamefont {Tanaka}}, \bibinfo {author} {\bibfnamefont
  {T.}~\bibnamefont {Katayama}}, \bibinfo {author} {\bibfnamefont
  {T.}~\bibnamefont {Togashi}}, \bibinfo {author} {\bibfnamefont
  {S.}~\bibnamefont {Owada}}, \bibinfo {author} {\bibfnamefont {Y.-i.}\
  \bibnamefont {Yamamoto}}, \bibinfo {author} {\bibfnamefont {S.}~\bibnamefont
  {Karashima}}, \bibinfo {author} {\bibfnamefont {J.}~\bibnamefont
  {Nishitani}}, \bibinfo {author} {\bibfnamefont {M.}~\bibnamefont {Yabashi}},
  \bibinfo {author} {\bibfnamefont {T.}~\bibnamefont {Suzuki}},\ and\ \bibinfo
  {author} {\bibfnamefont {K.}~\bibnamefont {Misawa}},\ }\bibfield  {title}
  {\bibinfo {title} {{Femtosecond time-resolved X-ray absorption spectroscopy
  of anatase TiO2 nanoparticles using XFEL}},\ }\href
  {https://doi.org/10.1063/1.4989862} {\bibfield  {journal} {\bibinfo
  {journal} {Structural Dynamics}\ }\textbf {\bibinfo {volume} {4}},\ \bibinfo
  {pages} {044033} (\bibinfo {year} {2017})}\BibitemShut {NoStop}%
\bibitem [{\citenamefont {G{\'{e}}neaux}\ \emph {et~al.}(2019)\citenamefont
  {G{\'{e}}neaux}, \citenamefont {Marroux}, \citenamefont {Guggenmos},
  \citenamefont {Neumark},\ and\ \citenamefont {Leone}}]{Geneaux2019}%
  \BibitemOpen
  \bibfield  {author} {\bibinfo {author} {\bibfnamefont {R.}~\bibnamefont
  {G{\'{e}}neaux}}, \bibinfo {author} {\bibfnamefont {H.~J.~B.}\ \bibnamefont
  {Marroux}}, \bibinfo {author} {\bibfnamefont {A.}~\bibnamefont {Guggenmos}},
  \bibinfo {author} {\bibfnamefont {D.~M.}\ \bibnamefont {Neumark}},\ and\
  \bibinfo {author} {\bibfnamefont {S.~R.}\ \bibnamefont {Leone}},\ }\bibfield
  {title} {\bibinfo {title} {{Transient absorption spectroscopy using high
  harmonic generation: a review of ultrafast X-ray dynamics in molecules and
  solids}},\ }\href {https://doi.org/10.1098/rsta.2017.0463} {\bibfield
  {journal} {\bibinfo  {journal} {Philosophical Transactions of the Royal
  Society A: Mathematical, Physical and Engineering Sciences}\ }\textbf
  {\bibinfo {volume} {377}},\ \bibinfo {pages} {20170463} (\bibinfo {year}
  {2019})}\BibitemShut {NoStop}%
\bibitem [{\citenamefont {Gallmann}\ \emph {et~al.}(2013)\citenamefont
  {Gallmann}, \citenamefont {Herrmann}, \citenamefont {Locher}, \citenamefont
  {Sabbar}, \citenamefont {Ludwig}, \citenamefont {Lucchini},\ and\
  \citenamefont {Keller}}]{Gallmann2013}%
  \BibitemOpen
  \bibfield  {author} {\bibinfo {author} {\bibfnamefont {L.}~\bibnamefont
  {Gallmann}}, \bibinfo {author} {\bibfnamefont {J.}~\bibnamefont {Herrmann}},
  \bibinfo {author} {\bibfnamefont {R.}~\bibnamefont {Locher}}, \bibinfo
  {author} {\bibfnamefont {M.}~\bibnamefont {Sabbar}}, \bibinfo {author}
  {\bibfnamefont {A.}~\bibnamefont {Ludwig}}, \bibinfo {author} {\bibfnamefont
  {M.}~\bibnamefont {Lucchini}},\ and\ \bibinfo {author} {\bibfnamefont
  {U.}~\bibnamefont {Keller}},\ }\bibfield  {title} {\bibinfo {title}
  {{Resolving intra-atomic electron dynamics with attosecond transient
  absorption spectroscopy}},\ }\href
  {https://doi.org/10.1080/00268976.2013.799298} {\bibfield  {journal}
  {\bibinfo  {journal} {Molecular Physics}\ }\textbf {\bibinfo {volume}
  {111}},\ \bibinfo {pages} {2243} (\bibinfo {year} {2013})}\BibitemShut
  {NoStop}%
\bibitem [{\citenamefont {Saito}\ \emph {et~al.}(2019)\citenamefont {Saito},
  \citenamefont {Sannohe}, \citenamefont {Ishii}, \citenamefont {Kanai},
  \citenamefont {Kosugi}, \citenamefont {Wu}, \citenamefont {Chew},
  \citenamefont {Han}, \citenamefont {Chang},\ and\ \citenamefont
  {Itatani}}]{Saito2019}%
  \BibitemOpen
  \bibfield  {author} {\bibinfo {author} {\bibfnamefont {N.}~\bibnamefont
  {Saito}}, \bibinfo {author} {\bibfnamefont {H.}~\bibnamefont {Sannohe}},
  \bibinfo {author} {\bibfnamefont {N.}~\bibnamefont {Ishii}}, \bibinfo
  {author} {\bibfnamefont {T.}~\bibnamefont {Kanai}}, \bibinfo {author}
  {\bibfnamefont {N.}~\bibnamefont {Kosugi}}, \bibinfo {author} {\bibfnamefont
  {Y.}~\bibnamefont {Wu}}, \bibinfo {author} {\bibfnamefont {A.}~\bibnamefont
  {Chew}}, \bibinfo {author} {\bibfnamefont {S.}~\bibnamefont {Han}}, \bibinfo
  {author} {\bibfnamefont {Z.}~\bibnamefont {Chang}},\ and\ \bibinfo {author}
  {\bibfnamefont {J.}~\bibnamefont {Itatani}},\ }\bibfield  {title} {\bibinfo
  {title} {{Real-time observation of electronic, vibrational, and rotational
  dynamics in nitric oxide with attosecond soft x-ray pulses at 400 eV}},\
  }\href {https://doi.org/10.1364/OPTICA.6.001542} {\bibfield  {journal}
  {\bibinfo  {journal} {Optica}\ }\textbf {\bibinfo {volume} {6}},\ \bibinfo
  {pages} {1542} (\bibinfo {year} {2019})},\ \Eprint
  {https://arxiv.org/abs/1904.10456} {arXiv:1904.10456} \BibitemShut {NoStop}%
\bibitem [{\citenamefont {Marroux}\ \emph {et~al.}(2020)\citenamefont
  {Marroux}, \citenamefont {Fidler}, \citenamefont {Ghosh}, \citenamefont
  {Kobayashi}, \citenamefont {Gokhberg}, \citenamefont {Kuleff}, \citenamefont
  {Leone},\ and\ \citenamefont {Neumark}}]{Marroux2020}%
  \BibitemOpen
  \bibfield  {author} {\bibinfo {author} {\bibfnamefont {H.~J.~B.}\
  \bibnamefont {Marroux}}, \bibinfo {author} {\bibfnamefont {A.~P.}\
  \bibnamefont {Fidler}}, \bibinfo {author} {\bibfnamefont {A.}~\bibnamefont
  {Ghosh}}, \bibinfo {author} {\bibfnamefont {Y.}~\bibnamefont {Kobayashi}},
  \bibinfo {author} {\bibfnamefont {K.}~\bibnamefont {Gokhberg}}, \bibinfo
  {author} {\bibfnamefont {A.~I.}\ \bibnamefont {Kuleff}}, \bibinfo {author}
  {\bibfnamefont {S.~R.}\ \bibnamefont {Leone}},\ and\ \bibinfo {author}
  {\bibfnamefont {D.~M.}\ \bibnamefont {Neumark}},\ }\bibfield  {title}
  {\bibinfo {title} {Attosecond spectroscopy reveals alignment dependent
  core-hole dynamics in the {ICl} molecule},\ }\href
  {https://doi.org/10.1038/s41467-020-19496-0} {\bibfield  {journal} {\bibinfo
  {journal} {Nature Communications}\ }\textbf {\bibinfo {volume} {11}},\
  \bibinfo {pages} {5810} (\bibinfo {year} {2020})}\BibitemShut {NoStop}%
\bibitem [{\citenamefont {Buades}\ \emph {et~al.}(2018)\citenamefont {Buades},
  \citenamefont {Pic{\'{o}}n}, \citenamefont {Le{\'{o}}n}, \citenamefont {{Di
  Palo}}, \citenamefont {Cousin}, \citenamefont {Cocchi}, \citenamefont
  {Pellegrin}, \citenamefont {Martin}, \citenamefont {Ma{\~{n}}as-Valero},
  \citenamefont {Coronado}, \citenamefont {Danz}, \citenamefont {Draxl},
  \citenamefont {Uemoto}, \citenamefont {Yabana}, \citenamefont {Schultze},
  \citenamefont {Wall},\ and\ \citenamefont {Biegert}}]{Buades2018}%
  \BibitemOpen
  \bibfield  {author} {\bibinfo {author} {\bibfnamefont {B.}~\bibnamefont
  {Buades}}, \bibinfo {author} {\bibfnamefont {A.}~\bibnamefont {Pic{\'{o}}n}},
  \bibinfo {author} {\bibfnamefont {I.}~\bibnamefont {Le{\'{o}}n}}, \bibinfo
  {author} {\bibfnamefont {N.}~\bibnamefont {{Di Palo}}}, \bibinfo {author}
  {\bibfnamefont {S.~L.}\ \bibnamefont {Cousin}}, \bibinfo {author}
  {\bibfnamefont {C.}~\bibnamefont {Cocchi}}, \bibinfo {author} {\bibfnamefont
  {E.}~\bibnamefont {Pellegrin}}, \bibinfo {author} {\bibfnamefont {J.~H.}\
  \bibnamefont {Martin}}, \bibinfo {author} {\bibfnamefont {S.}~\bibnamefont
  {Ma{\~{n}}as-Valero}}, \bibinfo {author} {\bibfnamefont {E.}~\bibnamefont
  {Coronado}}, \bibinfo {author} {\bibfnamefont {T.}~\bibnamefont {Danz}},
  \bibinfo {author} {\bibfnamefont {C.}~\bibnamefont {Draxl}}, \bibinfo
  {author} {\bibfnamefont {M.}~\bibnamefont {Uemoto}}, \bibinfo {author}
  {\bibfnamefont {K.}~\bibnamefont {Yabana}}, \bibinfo {author} {\bibfnamefont
  {M.}~\bibnamefont {Schultze}}, \bibinfo {author} {\bibfnamefont
  {S.}~\bibnamefont {Wall}},\ and\ \bibinfo {author} {\bibfnamefont
  {J.}~\bibnamefont {Biegert}},\ }\bibfield  {title} {\bibinfo {title}
  {{Attosecond-resolved petahertz carrier motion in semi-metallic TiS2}},\
  }\href {http://arxiv.org/abs/1808.06493} {\bibfield  {journal} {\bibinfo
  {journal} {arXiv e-prints, arXiv:1808.06493}\ } (\bibinfo {year} {2018})},\
  \Eprint {https://arxiv.org/abs/1808.06493} {arXiv:1808.06493} \BibitemShut
  {NoStop}%
\bibitem [{\citenamefont {Volkov}\ \emph
  {et~al.}(2019{\natexlab{a}})\citenamefont {Volkov}, \citenamefont {Sato},
  \citenamefont {Schlaepfer}, \citenamefont {Kasmi}, \citenamefont {Hartmann},
  \citenamefont {Lucchini}, \citenamefont {Gallmann}, \citenamefont {Rubio},\
  and\ \citenamefont {Keller}}]{Volkov2019a}%
  \BibitemOpen
  \bibfield  {author} {\bibinfo {author} {\bibfnamefont {M.}~\bibnamefont
  {Volkov}}, \bibinfo {author} {\bibfnamefont {S.~A.}\ \bibnamefont {Sato}},
  \bibinfo {author} {\bibfnamefont {F.}~\bibnamefont {Schlaepfer}}, \bibinfo
  {author} {\bibfnamefont {L.}~\bibnamefont {Kasmi}}, \bibinfo {author}
  {\bibfnamefont {N.}~\bibnamefont {Hartmann}}, \bibinfo {author}
  {\bibfnamefont {M.}~\bibnamefont {Lucchini}}, \bibinfo {author}
  {\bibfnamefont {L.}~\bibnamefont {Gallmann}}, \bibinfo {author}
  {\bibfnamefont {A.}~\bibnamefont {Rubio}},\ and\ \bibinfo {author}
  {\bibfnamefont {U.}~\bibnamefont {Keller}},\ }\bibfield  {title} {\bibinfo
  {title} {{Attosecond screening dynamics mediated by electron localization in
  transition metals}},\ }\href {https://doi.org/10.1038/s41567-019-0602-9}
  {\bibfield  {journal} {\bibinfo  {journal} {Nature Physics}\ }\textbf
  {\bibinfo {volume} {15}},\ \bibinfo {pages} {1145} (\bibinfo {year}
  {2019}{\natexlab{a}})}\BibitemShut {NoStop}%
\bibitem [{\citenamefont {Lang}(2018)}]{Lang2018}%
  \BibitemOpen
  \bibfield  {author} {\bibinfo {author} {\bibfnamefont {B.}~\bibnamefont
  {Lang}},\ }\bibfield  {title} {\bibinfo {title} {{Photometrics of ultrafast
  and fast broadband electronic transient absorption spectroscopy: State of the
  art}},\ }\href {https://doi.org/10.1063/1.5039457} {\bibfield  {journal}
  {\bibinfo  {journal} {Review of Scientific Instruments}\ }\textbf {\bibinfo
  {volume} {89}},\ \bibinfo {pages} {093112} (\bibinfo {year}
  {2018})}\BibitemShut {NoStop}%
\bibitem [{\citenamefont {Oppermann}\ \emph {et~al.}(2019)\citenamefont
  {Oppermann}, \citenamefont {Bauer}, \citenamefont {Rossi}, \citenamefont
  {Zinna}, \citenamefont {Helbing}, \citenamefont {Lacour},\ and\ \citenamefont
  {Chergui}}]{Oppermann2019}%
  \BibitemOpen
  \bibfield  {author} {\bibinfo {author} {\bibfnamefont {M.}~\bibnamefont
  {Oppermann}}, \bibinfo {author} {\bibfnamefont {B.}~\bibnamefont {Bauer}},
  \bibinfo {author} {\bibfnamefont {T.}~\bibnamefont {Rossi}}, \bibinfo
  {author} {\bibfnamefont {F.}~\bibnamefont {Zinna}}, \bibinfo {author}
  {\bibfnamefont {J.}~\bibnamefont {Helbing}}, \bibinfo {author} {\bibfnamefont
  {J.}~\bibnamefont {Lacour}},\ and\ \bibinfo {author} {\bibfnamefont
  {M.}~\bibnamefont {Chergui}},\ }\bibfield  {title} {\bibinfo {title}
  {{Ultrafast broadband circular dichroism in the deep ultraviolet}},\ }\href
  {https://doi.org/10.1364/OPTICA.6.000056} {\bibfield  {journal} {\bibinfo
  {journal} {Optica}\ }\textbf {\bibinfo {volume} {6}},\ \bibinfo {pages} {56}
  (\bibinfo {year} {2019})}\BibitemShut {NoStop}%
\bibitem [{\citenamefont {Volkov}\ \emph
  {et~al.}(2019{\natexlab{b}})\citenamefont {Volkov}, \citenamefont {Pupeikis},
  \citenamefont {Phillips}, \citenamefont {Schlaepfer}, \citenamefont
  {Gallmann},\ and\ \citenamefont {Keller}}]{Volkov2019}%
  \BibitemOpen
  \bibfield  {author} {\bibinfo {author} {\bibfnamefont {M.}~\bibnamefont
  {Volkov}}, \bibinfo {author} {\bibfnamefont {J.}~\bibnamefont {Pupeikis}},
  \bibinfo {author} {\bibfnamefont {C.~R.}\ \bibnamefont {Phillips}}, \bibinfo
  {author} {\bibfnamefont {F.}~\bibnamefont {Schlaepfer}}, \bibinfo {author}
  {\bibfnamefont {L.}~\bibnamefont {Gallmann}},\ and\ \bibinfo {author}
  {\bibfnamefont {U.}~\bibnamefont {Keller}},\ }\bibfield  {title} {\bibinfo
  {title} {{Reduction of laser-intensity-correlated noise in high-harmonic
  generation}},\ }\href {https://doi.org/10.1364/OE.27.007886} {\bibfield
  {journal} {\bibinfo  {journal} {Optics Express}\ }\textbf {\bibinfo {volume}
  {27}},\ \bibinfo {pages} {7886} (\bibinfo {year}
  {2019}{\natexlab{b}})}\BibitemShut {NoStop}%
\bibitem [{\citenamefont {Stoo{\ss}}\ \emph {et~al.}(2019)\citenamefont
  {Stoo{\ss}}, \citenamefont {Hartmann}, \citenamefont {Birk}, \citenamefont
  {Borisova}, \citenamefont {Ding}, \citenamefont {Bl{\"{a}}ttermann},
  \citenamefont {Ott},\ and\ \citenamefont {Pfeifer}}]{Stooss2019}%
  \BibitemOpen
  \bibfield  {author} {\bibinfo {author} {\bibfnamefont {V.}~\bibnamefont
  {Stoo{\ss}}}, \bibinfo {author} {\bibfnamefont {M.}~\bibnamefont {Hartmann}},
  \bibinfo {author} {\bibfnamefont {P.}~\bibnamefont {Birk}}, \bibinfo {author}
  {\bibfnamefont {G.~D.}\ \bibnamefont {Borisova}}, \bibinfo {author}
  {\bibfnamefont {T.}~\bibnamefont {Ding}}, \bibinfo {author} {\bibfnamefont
  {A.}~\bibnamefont {Bl{\"{a}}ttermann}}, \bibinfo {author} {\bibfnamefont
  {C.}~\bibnamefont {Ott}},\ and\ \bibinfo {author} {\bibfnamefont
  {T.}~\bibnamefont {Pfeifer}},\ }\bibfield  {title} {\bibinfo {title}
  {{XUV-beamline for attosecond transient absorption measurements featuring a
  broadband common beam-path time-delay unit and in situ reference spectrometer
  for high stability and sensitivity}},\ }\href
  {https://doi.org/10.1063/1.5091069} {\bibfield  {journal} {\bibinfo
  {journal} {Review of Scientific Instruments}\ }\textbf {\bibinfo {volume}
  {90}},\ \bibinfo {pages} {053108} (\bibinfo {year} {2019})}\BibitemShut
  {NoStop}%
\bibitem [{\citenamefont {Willems}\ \emph {et~al.}(2020)\citenamefont
  {Willems}, \citenamefont {{von Korff Schmising}}, \citenamefont
  {Str{\"{u}}ber}, \citenamefont {Schick}, \citenamefont {Engel}, \citenamefont
  {Dewhurst}, \citenamefont {Elliott}, \citenamefont {Sharma},\ and\
  \citenamefont {Eisebitt}}]{Willems2020}%
  \BibitemOpen
  \bibfield  {author} {\bibinfo {author} {\bibfnamefont {F.}~\bibnamefont
  {Willems}}, \bibinfo {author} {\bibfnamefont {C.}~\bibnamefont {{von Korff
  Schmising}}}, \bibinfo {author} {\bibfnamefont {C.}~\bibnamefont
  {Str{\"{u}}ber}}, \bibinfo {author} {\bibfnamefont {D.}~\bibnamefont
  {Schick}}, \bibinfo {author} {\bibfnamefont {D.~W.}\ \bibnamefont {Engel}},
  \bibinfo {author} {\bibfnamefont {J.~K.}\ \bibnamefont {Dewhurst}}, \bibinfo
  {author} {\bibfnamefont {P.}~\bibnamefont {Elliott}}, \bibinfo {author}
  {\bibfnamefont {S.}~\bibnamefont {Sharma}},\ and\ \bibinfo {author}
  {\bibfnamefont {S.}~\bibnamefont {Eisebitt}},\ }\bibfield  {title} {\bibinfo
  {title} {{Optical inter-site spin transfer probed by energy and spin-resolved
  transient absorption spectroscopy}},\ }\href
  {https://doi.org/10.1038/s41467-020-14691-5} {\bibfield  {journal} {\bibinfo
  {journal} {Nature Communications}\ }\textbf {\bibinfo {volume} {11}},\
  \bibinfo {pages} {871} (\bibinfo {year} {2020})}\BibitemShut {NoStop}%
\bibitem [{\citenamefont {Feng}\ \emph {et~al.}(2017)\citenamefont {Feng},
  \citenamefont {Vinogradov},\ and\ \citenamefont {Ge}}]{Feng2017}%
  \BibitemOpen
  \bibfield  {author} {\bibinfo {author} {\bibfnamefont {Y.}~\bibnamefont
  {Feng}}, \bibinfo {author} {\bibfnamefont {I.}~\bibnamefont {Vinogradov}},\
  and\ \bibinfo {author} {\bibfnamefont {N.-H.}\ \bibnamefont {Ge}},\
  }\bibfield  {title} {\bibinfo {title} {{General noise suppression scheme with
  reference detection in heterodyne nonlinear spectroscopy}},\ }\href
  {https://doi.org/10.1364/OE.25.026262} {\bibfield  {journal} {\bibinfo
  {journal} {Optics Express}\ }\textbf {\bibinfo {volume} {25}},\ \bibinfo
  {pages} {26262} (\bibinfo {year} {2017})}\BibitemShut {NoStop}%
\bibitem [{\citenamefont {Robben}\ and\ \citenamefont
  {Cheatum}(2020)}]{Robben2020}%
  \BibitemOpen
  \bibfield  {author} {\bibinfo {author} {\bibfnamefont {K.~C.}\ \bibnamefont
  {Robben}}\ and\ \bibinfo {author} {\bibfnamefont {C.~M.}\ \bibnamefont
  {Cheatum}},\ }\bibfield  {title} {\bibinfo {title} {{Edge-pixel referencing
  suppresses correlated baseline noise in heterodyned spectroscopies}},\ }\href
  {https://doi.org/10.1063/1.5134987} {\bibfield  {journal} {\bibinfo
  {journal} {The Journal of Chemical Physics}\ }\textbf {\bibinfo {volume}
  {152}},\ \bibinfo {pages} {094201} (\bibinfo {year} {2020})}\BibitemShut
  {NoStop}%
\bibitem [{\citenamefont {Erny}\ \emph {et~al.}(2011)\citenamefont {Erny},
  \citenamefont {Mansten}, \citenamefont {Gisselbrecht}, \citenamefont
  {Schwenke}, \citenamefont {Rakowski}, \citenamefont {He}, \citenamefont
  {Gaarde}, \citenamefont {Werin},\ and\ \citenamefont
  {L'Huillier}}]{Erny2011}%
  \BibitemOpen
  \bibfield  {author} {\bibinfo {author} {\bibfnamefont {C.}~\bibnamefont
  {Erny}}, \bibinfo {author} {\bibfnamefont {E.}~\bibnamefont {Mansten}},
  \bibinfo {author} {\bibfnamefont {M.}~\bibnamefont {Gisselbrecht}}, \bibinfo
  {author} {\bibfnamefont {J.}~\bibnamefont {Schwenke}}, \bibinfo {author}
  {\bibfnamefont {R.}~\bibnamefont {Rakowski}}, \bibinfo {author}
  {\bibfnamefont {X.}~\bibnamefont {He}}, \bibinfo {author} {\bibfnamefont
  {M.~B.}\ \bibnamefont {Gaarde}}, \bibinfo {author} {\bibfnamefont
  {S.}~\bibnamefont {Werin}},\ and\ \bibinfo {author} {\bibfnamefont
  {A.}~\bibnamefont {L'Huillier}},\ }\bibfield  {title} {\bibinfo {title}
  {{Metrology of high-order harmonics for free-electron laser seeding}},\
  }\href {https://doi.org/10.1088/1367-2630/13/7/073035} {\bibfield  {journal}
  {\bibinfo  {journal} {New Journal of Physics}\ }\textbf {\bibinfo {volume}
  {13}},\ \bibinfo {pages} {073035} (\bibinfo {year} {2011})}\BibitemShut
  {NoStop}%
\bibitem [{\citenamefont {Kastl}\ \emph {et~al.}(2017)\citenamefont {Kastl},
  \citenamefont {Chen}, \citenamefont {Kuykendall}, \citenamefont {Shevitski},
  \citenamefont {Darlington}, \citenamefont {Borys}, \citenamefont {Krayev},
  \citenamefont {Schuck}, \citenamefont {Aloni},\ and\ \citenamefont
  {Schwartzberg}}]{Kastl2017}%
  \BibitemOpen
  \bibfield  {author} {\bibinfo {author} {\bibfnamefont {C.}~\bibnamefont
  {Kastl}}, \bibinfo {author} {\bibfnamefont {C.~T.}\ \bibnamefont {Chen}},
  \bibinfo {author} {\bibfnamefont {T.}~\bibnamefont {Kuykendall}}, \bibinfo
  {author} {\bibfnamefont {B.}~\bibnamefont {Shevitski}}, \bibinfo {author}
  {\bibfnamefont {T.~P.}\ \bibnamefont {Darlington}}, \bibinfo {author}
  {\bibfnamefont {N.~J.}\ \bibnamefont {Borys}}, \bibinfo {author}
  {\bibfnamefont {A.}~\bibnamefont {Krayev}}, \bibinfo {author} {\bibfnamefont
  {P.~J.}\ \bibnamefont {Schuck}}, \bibinfo {author} {\bibfnamefont
  {S.}~\bibnamefont {Aloni}},\ and\ \bibinfo {author} {\bibfnamefont {A.~M.}\
  \bibnamefont {Schwartzberg}},\ }\bibfield  {title} {\bibinfo {title} {The
  important role of water in growth of monolayer transition metal
  dichalcogenides},\ }\href {https://doi.org/10.1088/2053-1583/aa5f4d}
  {\bibfield  {journal} {\bibinfo  {journal} {2D Mater.}\ }\textbf {\bibinfo
  {volume} {4}},\ \bibinfo {pages} {021024} (\bibinfo {year}
  {2017})}\BibitemShut {NoStop}%
\bibitem [{\citenamefont {Kaplan}\ \emph {et~al.}(2018)\citenamefont {Kaplan},
  \citenamefont {Kraus}, \citenamefont {Ross}, \citenamefont {Z{\"{u}}rch},
  \citenamefont {Cushing}, \citenamefont {Jager}, \citenamefont {Chang},
  \citenamefont {Gullikson}, \citenamefont {Neumark},\ and\ \citenamefont
  {Leone}}]{Kaplan2018}%
  \BibitemOpen
  \bibfield  {author} {\bibinfo {author} {\bibfnamefont {C.~J.}\ \bibnamefont
  {Kaplan}}, \bibinfo {author} {\bibfnamefont {P.~M.}\ \bibnamefont {Kraus}},
  \bibinfo {author} {\bibfnamefont {A.~D.}\ \bibnamefont {Ross}}, \bibinfo
  {author} {\bibfnamefont {M.}~\bibnamefont {Z{\"{u}}rch}}, \bibinfo {author}
  {\bibfnamefont {S.~K.}\ \bibnamefont {Cushing}}, \bibinfo {author}
  {\bibfnamefont {M.~F.}\ \bibnamefont {Jager}}, \bibinfo {author}
  {\bibfnamefont {H.-T.}\ \bibnamefont {Chang}}, \bibinfo {author}
  {\bibfnamefont {E.~M.}\ \bibnamefont {Gullikson}}, \bibinfo {author}
  {\bibfnamefont {D.~M.}\ \bibnamefont {Neumark}},\ and\ \bibinfo {author}
  {\bibfnamefont {S.~R.}\ \bibnamefont {Leone}},\ }\bibfield  {title} {\bibinfo
  {title} {{Femtosecond tracking of carrier relaxation in germanium with
  extreme ultraviolet transient reflectivity}},\ }\href
  {https://doi.org/10.1103/PhysRevB.97.205202} {\bibfield  {journal} {\bibinfo
  {journal} {Physical Review B}\ }\textbf {\bibinfo {volume} {97}},\ \bibinfo
  {pages} {205202} (\bibinfo {year} {2018})}\BibitemShut {NoStop}%
\bibitem [{\citenamefont {Moulet}\ \emph {et~al.}(2017)\citenamefont {Moulet},
  \citenamefont {Bertrand}, \citenamefont {Klostermann}, \citenamefont
  {Guggenmos}, \citenamefont {Karpowicz},\ and\ \citenamefont
  {Goulielmakis}}]{Moulet2017}%
  \BibitemOpen
  \bibfield  {author} {\bibinfo {author} {\bibfnamefont {A.}~\bibnamefont
  {Moulet}}, \bibinfo {author} {\bibfnamefont {J.~B.}\ \bibnamefont
  {Bertrand}}, \bibinfo {author} {\bibfnamefont {T.}~\bibnamefont
  {Klostermann}}, \bibinfo {author} {\bibfnamefont {A.}~\bibnamefont
  {Guggenmos}}, \bibinfo {author} {\bibfnamefont {N.}~\bibnamefont
  {Karpowicz}},\ and\ \bibinfo {author} {\bibfnamefont {E.}~\bibnamefont
  {Goulielmakis}},\ }\bibfield  {title} {\bibinfo {title} {{Soft x-ray
  excitonics}},\ }\href {https://doi.org/10.1126/science.aan4737} {\bibfield
  {journal} {\bibinfo  {journal} {Science}\ }\textbf {\bibinfo {volume}
  {357}},\ \bibinfo {pages} {1134} (\bibinfo {year} {2017})}\BibitemShut
  {NoStop}%
\bibitem [{\citenamefont {G{\'{e}}neaux}\ \emph {et~al.}(2020)\citenamefont
  {G{\'{e}}neaux}, \citenamefont {Kaplan}, \citenamefont {Yue}, \citenamefont
  {Ross}, \citenamefont {B{\ae}kh{\o}j}, \citenamefont {Kraus}, \citenamefont
  {Chang}, \citenamefont {Guggenmos}, \citenamefont {Huang}, \citenamefont
  {Z{\"{u}}rch}, \citenamefont {Schafer}, \citenamefont {Neumark},
  \citenamefont {Gaarde},\ and\ \citenamefont {Leone}}]{Geneaux2020}%
  \BibitemOpen
  \bibfield  {author} {\bibinfo {author} {\bibfnamefont {R.}~\bibnamefont
  {G{\'{e}}neaux}}, \bibinfo {author} {\bibfnamefont {C.~J.}\ \bibnamefont
  {Kaplan}}, \bibinfo {author} {\bibfnamefont {L.}~\bibnamefont {Yue}},
  \bibinfo {author} {\bibfnamefont {A.~D.}\ \bibnamefont {Ross}}, \bibinfo
  {author} {\bibfnamefont {J.~E.}\ \bibnamefont {B{\ae}kh{\o}j}}, \bibinfo
  {author} {\bibfnamefont {P.~M.}\ \bibnamefont {Kraus}}, \bibinfo {author}
  {\bibfnamefont {H.-T.}\ \bibnamefont {Chang}}, \bibinfo {author}
  {\bibfnamefont {A.}~\bibnamefont {Guggenmos}}, \bibinfo {author}
  {\bibfnamefont {M.-Y.}\ \bibnamefont {Huang}}, \bibinfo {author}
  {\bibfnamefont {M.}~\bibnamefont {Z{\"{u}}rch}}, \bibinfo {author}
  {\bibfnamefont {K.~J.}\ \bibnamefont {Schafer}}, \bibinfo {author}
  {\bibfnamefont {D.~M.}\ \bibnamefont {Neumark}}, \bibinfo {author}
  {\bibfnamefont {M.~B.}\ \bibnamefont {Gaarde}},\ and\ \bibinfo {author}
  {\bibfnamefont {S.~R.}\ \bibnamefont {Leone}},\ }\bibfield  {title} {\bibinfo
  {title} {{Attosecond Time-Domain Measurement of Core-Level-Exciton Decay in
  Magnesium Oxide}},\ }\href {https://doi.org/10.1103/PhysRevLett.124.207401}
  {\bibfield  {journal} {\bibinfo  {journal} {Physical Review Letters}\
  }\textbf {\bibinfo {volume} {124}},\ \bibinfo {pages} {207401} (\bibinfo
  {year} {2020})}\BibitemShut {NoStop}%
\bibitem [{\citenamefont {Lucchini}\ \emph {et~al.}(2020)\citenamefont
  {Lucchini}, \citenamefont {Sato}, \citenamefont {Lucarelli}, \citenamefont
  {Moio}, \citenamefont {Inzani}, \citenamefont {Borrego-Varillas},
  \citenamefont {Frassetto}, \citenamefont {Poletto}, \citenamefont
  {H{\"{u}}bener}, \citenamefont {Giovannini}, \citenamefont {Rubio},\ and\
  \citenamefont {Nisoli}}]{lucchini2020unravelling}%
  \BibitemOpen
  \bibfield  {author} {\bibinfo {author} {\bibfnamefont {M.}~\bibnamefont
  {Lucchini}}, \bibinfo {author} {\bibfnamefont {S.~A.}\ \bibnamefont {Sato}},
  \bibinfo {author} {\bibfnamefont {G.~D.}\ \bibnamefont {Lucarelli}}, \bibinfo
  {author} {\bibfnamefont {B.}~\bibnamefont {Moio}}, \bibinfo {author}
  {\bibfnamefont {G.}~\bibnamefont {Inzani}}, \bibinfo {author} {\bibfnamefont
  {R.}~\bibnamefont {Borrego-Varillas}}, \bibinfo {author} {\bibfnamefont
  {F.}~\bibnamefont {Frassetto}}, \bibinfo {author} {\bibfnamefont
  {L.}~\bibnamefont {Poletto}}, \bibinfo {author} {\bibfnamefont
  {H.}~\bibnamefont {H{\"{u}}bener}}, \bibinfo {author} {\bibfnamefont {U.~D.}\
  \bibnamefont {Giovannini}}, \bibinfo {author} {\bibfnamefont
  {A.}~\bibnamefont {Rubio}},\ and\ \bibinfo {author} {\bibfnamefont
  {M.}~\bibnamefont {Nisoli}},\ }\bibfield  {title} {\bibinfo {title}
  {{Unravelling the intertwined atomic and bulk nature of localised excitons by
  attosecond spectroscopy}},\ }\href@noop {} {\bibfield  {journal} {\bibinfo
  {journal} {arXiv e-prints, arXiv:2006.16008}\ } (\bibinfo {year} {2020})},\
  \Eprint {https://arxiv.org/abs/2006.16008} {arXiv:2006.16008
  [physics.optics]} \BibitemShut {NoStop}%
\bibitem [{\citenamefont {Kuc}\ \emph {et~al.}(2011)\citenamefont {Kuc},
  \citenamefont {Zibouche},\ and\ \citenamefont {Heine}}]{Kuc2011}%
  \BibitemOpen
  \bibfield  {author} {\bibinfo {author} {\bibfnamefont {A.}~\bibnamefont
  {Kuc}}, \bibinfo {author} {\bibfnamefont {N.}~\bibnamefont {Zibouche}},\ and\
  \bibinfo {author} {\bibfnamefont {T.}~\bibnamefont {Heine}},\ }\bibfield
  {title} {\bibinfo {title} {{Influence of quantum confinement on the
  electronic structure of the transition metal sulfide TS2}},\ }\href
  {https://doi.org/10.1103/PhysRevB.83.245213} {\bibfield  {journal} {\bibinfo
  {journal} {Physical Review B}\ }\textbf {\bibinfo {volume} {83}},\ \bibinfo
  {pages} {245213} (\bibinfo {year} {2011})}\BibitemShut {NoStop}%
\bibitem [{\citenamefont {Z{\"{u}}rch}\ \emph {et~al.}(2017)\citenamefont
  {Z{\"{u}}rch}, \citenamefont {Chang}, \citenamefont {Kraus}, \citenamefont
  {Cushing}, \citenamefont {Borja}, \citenamefont {Gandman}, \citenamefont
  {Kaplan}, \citenamefont {Oh}, \citenamefont {Prell}, \citenamefont
  {Prendergast}, \citenamefont {Pemmaraju}, \citenamefont {Neumark},\ and\
  \citenamefont {Leone}}]{Zurch2017}%
  \BibitemOpen
  \bibfield  {author} {\bibinfo {author} {\bibfnamefont {M.}~\bibnamefont
  {Z{\"{u}}rch}}, \bibinfo {author} {\bibfnamefont {H.-T.}\ \bibnamefont
  {Chang}}, \bibinfo {author} {\bibfnamefont {P.~M.}\ \bibnamefont {Kraus}},
  \bibinfo {author} {\bibfnamefont {S.~K.}\ \bibnamefont {Cushing}}, \bibinfo
  {author} {\bibfnamefont {L.~J.}\ \bibnamefont {Borja}}, \bibinfo {author}
  {\bibfnamefont {A.}~\bibnamefont {Gandman}}, \bibinfo {author} {\bibfnamefont
  {C.~J.}\ \bibnamefont {Kaplan}}, \bibinfo {author} {\bibfnamefont {M.~H.}\
  \bibnamefont {Oh}}, \bibinfo {author} {\bibfnamefont {J.~S.}\ \bibnamefont
  {Prell}}, \bibinfo {author} {\bibfnamefont {D.}~\bibnamefont {Prendergast}},
  \bibinfo {author} {\bibfnamefont {C.~D.}\ \bibnamefont {Pemmaraju}}, \bibinfo
  {author} {\bibfnamefont {D.~M.}\ \bibnamefont {Neumark}},\ and\ \bibinfo
  {author} {\bibfnamefont {S.~R.}\ \bibnamefont {Leone}},\ }\bibfield  {title}
  {\bibinfo {title} {{Ultrafast carrier thermalization and trapping in
  silicon-germanium alloy probed by extreme ultraviolet transient absorption
  spectroscopy}},\ }\href {https://doi.org/10.1063/1.4985056} {\bibfield
  {journal} {\bibinfo  {journal} {Structural Dynamics}\ }\textbf {\bibinfo
  {volume} {4}},\ \bibinfo {pages} {044029} (\bibinfo {year}
  {2017})}\BibitemShut {NoStop}%
\bibitem [{\citenamefont {Ge}\ \emph {et~al.}(2013)\citenamefont {Ge},
  \citenamefont {Boutu}, \citenamefont {Gauthier}, \citenamefont {Wang},
  \citenamefont {Borta}, \citenamefont {Barbrel}, \citenamefont {Ducousso},
  \citenamefont {Gonzalez}, \citenamefont {Carr{\'{e}}}, \citenamefont
  {Guillaumet}, \citenamefont {Perdrix}, \citenamefont {Gobert}, \citenamefont
  {Gautier}, \citenamefont {Lambert}, \citenamefont {Maia}, \citenamefont
  {Hajdu}, \citenamefont {Zeitoun},\ and\ \citenamefont {Merdji}}]{GeOE2013}%
  \BibitemOpen
  \bibfield  {author} {\bibinfo {author} {\bibfnamefont {X.}~\bibnamefont
  {Ge}}, \bibinfo {author} {\bibfnamefont {W.}~\bibnamefont {Boutu}}, \bibinfo
  {author} {\bibfnamefont {D.}~\bibnamefont {Gauthier}}, \bibinfo {author}
  {\bibfnamefont {F.}~\bibnamefont {Wang}}, \bibinfo {author} {\bibfnamefont
  {A.}~\bibnamefont {Borta}}, \bibinfo {author} {\bibfnamefont
  {B.}~\bibnamefont {Barbrel}}, \bibinfo {author} {\bibfnamefont
  {M.}~\bibnamefont {Ducousso}}, \bibinfo {author} {\bibfnamefont {A.~I.}\
  \bibnamefont {Gonzalez}}, \bibinfo {author} {\bibfnamefont {B.}~\bibnamefont
  {Carr{\'{e}}}}, \bibinfo {author} {\bibfnamefont {D.}~\bibnamefont
  {Guillaumet}}, \bibinfo {author} {\bibfnamefont {M.}~\bibnamefont {Perdrix}},
  \bibinfo {author} {\bibfnamefont {O.}~\bibnamefont {Gobert}}, \bibinfo
  {author} {\bibfnamefont {J.}~\bibnamefont {Gautier}}, \bibinfo {author}
  {\bibfnamefont {G.}~\bibnamefont {Lambert}}, \bibinfo {author} {\bibfnamefont
  {F.~R. N.~C.}\ \bibnamefont {Maia}}, \bibinfo {author} {\bibfnamefont
  {J.}~\bibnamefont {Hajdu}}, \bibinfo {author} {\bibfnamefont
  {P.}~\bibnamefont {Zeitoun}},\ and\ \bibinfo {author} {\bibfnamefont
  {H.}~\bibnamefont {Merdji}},\ }\bibfield  {title} {\bibinfo {title} {{Impact
  of wave front and coherence optimization in coherent diffractive imaging}},\
  }\href {https://doi.org/10.1364/OE.21.011441} {\bibfield  {journal} {\bibinfo
   {journal} {Opt. Express}\ }\textbf {\bibinfo {volume} {21}},\ \bibinfo
  {pages} {11441} (\bibinfo {year} {2013})}\BibitemShut {NoStop}%
\bibitem [{\citenamefont {G\'{e}neaux}(2020)}]{github-edge-referencing}%
  \BibitemOpen
  \bibfield  {author} {\bibinfo {author} {\bibfnamefont {R.}~\bibnamefont
  {G\'{e}neaux}},\ }\href@noop {} {}\bibinfo {howpublished}
  {\url{https://github.com/rgeneaux/edge-pixel-referencing}} (\bibinfo {year}
  {2020})\BibitemShut {NoStop}%
\bibitem [{\citenamefont {Kubin}\ \emph {et~al.}(2018)\citenamefont {Kubin},
  \citenamefont {Guo}, \citenamefont {Ekimova}, \citenamefont {Baker},
  \citenamefont {Kroll}, \citenamefont {K{\"{a}}llman}, \citenamefont {Kern},
  \citenamefont {Yachandra}, \citenamefont {Yano}, \citenamefont {Nibbering},
  \citenamefont {Lundberg},\ and\ \citenamefont {Wernet}}]{Kubin2018}%
  \BibitemOpen
  \bibfield  {author} {\bibinfo {author} {\bibfnamefont {M.}~\bibnamefont
  {Kubin}}, \bibinfo {author} {\bibfnamefont {M.}~\bibnamefont {Guo}}, \bibinfo
  {author} {\bibfnamefont {M.}~\bibnamefont {Ekimova}}, \bibinfo {author}
  {\bibfnamefont {M.~L.}\ \bibnamefont {Baker}}, \bibinfo {author}
  {\bibfnamefont {T.}~\bibnamefont {Kroll}}, \bibinfo {author} {\bibfnamefont
  {E.}~\bibnamefont {K{\"{a}}llman}}, \bibinfo {author} {\bibfnamefont
  {J.}~\bibnamefont {Kern}}, \bibinfo {author} {\bibfnamefont {V.~K.}\
  \bibnamefont {Yachandra}}, \bibinfo {author} {\bibfnamefont {J.}~\bibnamefont
  {Yano}}, \bibinfo {author} {\bibfnamefont {E.~T.~J.}\ \bibnamefont
  {Nibbering}}, \bibinfo {author} {\bibfnamefont {M.}~\bibnamefont
  {Lundberg}},\ and\ \bibinfo {author} {\bibfnamefont {P.}~\bibnamefont
  {Wernet}},\ }\bibfield  {title} {\bibinfo {title} {{Direct Determination of
  Absolute Absorption Cross Sections at the L-Edge of Dilute Mn Complexes in
  Solution Using a Transmission Flatjet}},\ }\href
  {https://doi.org/10.1021/acs.inorgchem.8b00419} {\bibfield  {journal}
  {\bibinfo  {journal} {Inorganic Chemistry}\ }\textbf {\bibinfo {volume}
  {57}},\ \bibinfo {pages} {5449} (\bibinfo {year} {2018})}\BibitemShut
  {NoStop}%
\bibitem [{\citenamefont {Kleine}\ \emph {et~al.}(2019)\citenamefont {Kleine},
  \citenamefont {Ekimova}, \citenamefont {Goldsztejn}, \citenamefont {Raabe},
  \citenamefont {Str{\"{u}}ber}, \citenamefont {Ludwig}, \citenamefont
  {Yarlagadda}, \citenamefont {Eisebitt}, \citenamefont {Vrakking},
  \citenamefont {Elsaesser}, \citenamefont {Nibbering},\ and\ \citenamefont
  {Rouz{\'{e}}e}}]{Kleine2019}%
  \BibitemOpen
  \bibfield  {author} {\bibinfo {author} {\bibfnamefont {C.}~\bibnamefont
  {Kleine}}, \bibinfo {author} {\bibfnamefont {M.}~\bibnamefont {Ekimova}},
  \bibinfo {author} {\bibfnamefont {G.}~\bibnamefont {Goldsztejn}}, \bibinfo
  {author} {\bibfnamefont {S.}~\bibnamefont {Raabe}}, \bibinfo {author}
  {\bibfnamefont {C.}~\bibnamefont {Str{\"{u}}ber}}, \bibinfo {author}
  {\bibfnamefont {J.}~\bibnamefont {Ludwig}}, \bibinfo {author} {\bibfnamefont
  {S.}~\bibnamefont {Yarlagadda}}, \bibinfo {author} {\bibfnamefont
  {S.}~\bibnamefont {Eisebitt}}, \bibinfo {author} {\bibfnamefont {M.~J.}\
  \bibnamefont {Vrakking}}, \bibinfo {author} {\bibfnamefont {T.}~\bibnamefont
  {Elsaesser}}, \bibinfo {author} {\bibfnamefont {E.~T.}\ \bibnamefont
  {Nibbering}},\ and\ \bibinfo {author} {\bibfnamefont {A.}~\bibnamefont
  {Rouz{\'{e}}e}},\ }\bibfield  {title} {\bibinfo {title} {{Soft X-ray
  Absorption Spectroscopy of Aqueous Solutions Using a Table-Top Femtosecond
  Soft X-ray Source}},\ }\href {https://doi.org/10.1021/acs.jpclett.8b03420}
  {\bibfield  {journal} {\bibinfo  {journal} {Journal of Physical Chemistry
  Letters}\ }\textbf {\bibinfo {volume} {10}},\ \bibinfo {pages} {52} (\bibinfo
  {year} {2019})}\BibitemShut {NoStop}%
\bibitem [{\citenamefont {Popmintchev}\ \emph {et~al.}(2012)\citenamefont
  {Popmintchev}, \citenamefont {Chen}, \citenamefont {Popmintchev},
  \citenamefont {Arpin}, \citenamefont {Brown}, \citenamefont
  {Ali{\v{s}}auskas}, \citenamefont {Andriukaitis}, \citenamefont
  {Bal{\v{c}}iunas}, \citenamefont {M{\"{u}}cke}, \citenamefont {Pugzlys},
  \citenamefont {Baltu{\v{s}}ka}, \citenamefont {Shim}, \citenamefont
  {Schrauth}, \citenamefont {Gaeta}, \citenamefont
  {Hern{\'{a}}ndez-Garc{\'{i}}a}, \citenamefont {Plaja}, \citenamefont
  {Becker}, \citenamefont {Jaron-Becker}, \citenamefont {Murnane},\ and\
  \citenamefont {Kapteyn}}]{Popmintchev2012b}%
  \BibitemOpen
  \bibfield  {author} {\bibinfo {author} {\bibfnamefont {T.}~\bibnamefont
  {Popmintchev}}, \bibinfo {author} {\bibfnamefont {M.~C.}\ \bibnamefont
  {Chen}}, \bibinfo {author} {\bibfnamefont {D.}~\bibnamefont {Popmintchev}},
  \bibinfo {author} {\bibfnamefont {P.}~\bibnamefont {Arpin}}, \bibinfo
  {author} {\bibfnamefont {S.}~\bibnamefont {Brown}}, \bibinfo {author}
  {\bibfnamefont {S.}~\bibnamefont {Ali{\v{s}}auskas}}, \bibinfo {author}
  {\bibfnamefont {G.}~\bibnamefont {Andriukaitis}}, \bibinfo {author}
  {\bibfnamefont {T.}~\bibnamefont {Bal{\v{c}}iunas}}, \bibinfo {author}
  {\bibfnamefont {O.~D.}\ \bibnamefont {M{\"{u}}cke}}, \bibinfo {author}
  {\bibfnamefont {A.}~\bibnamefont {Pugzlys}}, \bibinfo {author} {\bibfnamefont
  {A.}~\bibnamefont {Baltu{\v{s}}ka}}, \bibinfo {author} {\bibfnamefont
  {B.}~\bibnamefont {Shim}}, \bibinfo {author} {\bibfnamefont {S.~E.}\
  \bibnamefont {Schrauth}}, \bibinfo {author} {\bibfnamefont {A.}~\bibnamefont
  {Gaeta}}, \bibinfo {author} {\bibfnamefont {C.}~\bibnamefont
  {Hern{\'{a}}ndez-Garc{\'{i}}a}}, \bibinfo {author} {\bibfnamefont
  {L.}~\bibnamefont {Plaja}}, \bibinfo {author} {\bibfnamefont
  {A.}~\bibnamefont {Becker}}, \bibinfo {author} {\bibfnamefont
  {A.}~\bibnamefont {Jaron-Becker}}, \bibinfo {author} {\bibfnamefont {M.~M.}\
  \bibnamefont {Murnane}},\ and\ \bibinfo {author} {\bibfnamefont {H.~C.}\
  \bibnamefont {Kapteyn}},\ }\bibfield  {title} {\bibinfo {title} {{Bright
  coherent ultrahigh harmonics in the kev x-ray regime from mid-infrared
  femtosecond lasers}},\ }\href {https://doi.org/10.1126/science.1218497}
  {\bibfield  {journal} {\bibinfo  {journal} {Science}\ }\textbf {\bibinfo
  {volume} {336}},\ \bibinfo {pages} {1287} (\bibinfo {year}
  {2012})}\BibitemShut {NoStop}%
\bibitem [{\citenamefont {Johnson}\ \emph {et~al.}(2018)\citenamefont
  {Johnson}, \citenamefont {Austin}, \citenamefont {Wood}, \citenamefont
  {Brahms}, \citenamefont {Gregory}, \citenamefont {Holzner}, \citenamefont
  {Jarosch}, \citenamefont {Larsen}, \citenamefont {Parker}, \citenamefont
  {Str{\"{u}}ber}, \citenamefont {Ye}, \citenamefont {Tisch},\ and\
  \citenamefont {Marangos}}]{Johnson2018a}%
  \BibitemOpen
  \bibfield  {author} {\bibinfo {author} {\bibfnamefont {A.~S.}\ \bibnamefont
  {Johnson}}, \bibinfo {author} {\bibfnamefont {D.~R.}\ \bibnamefont {Austin}},
  \bibinfo {author} {\bibfnamefont {D.~A.}\ \bibnamefont {Wood}}, \bibinfo
  {author} {\bibfnamefont {C.}~\bibnamefont {Brahms}}, \bibinfo {author}
  {\bibfnamefont {A.}~\bibnamefont {Gregory}}, \bibinfo {author} {\bibfnamefont
  {K.~B.}\ \bibnamefont {Holzner}}, \bibinfo {author} {\bibfnamefont
  {S.}~\bibnamefont {Jarosch}}, \bibinfo {author} {\bibfnamefont {E.~W.}\
  \bibnamefont {Larsen}}, \bibinfo {author} {\bibfnamefont {S.}~\bibnamefont
  {Parker}}, \bibinfo {author} {\bibfnamefont {C.~S.}\ \bibnamefont
  {Str{\"{u}}ber}}, \bibinfo {author} {\bibfnamefont {P.}~\bibnamefont {Ye}},
  \bibinfo {author} {\bibfnamefont {J.~W.}\ \bibnamefont {Tisch}},\ and\
  \bibinfo {author} {\bibfnamefont {J.~P.}\ \bibnamefont {Marangos}},\
  }\bibfield  {title} {\bibinfo {title} {{High-flux soft x-ray harmonic
  generation from ionization-shaped few-cycle laser pulses}},\ }\bibfield
  {journal} {\bibinfo  {journal} {Science Advances}\ }\textbf {\bibinfo
  {volume} {4}},\ \href {https://doi.org/10.1126/sciadv.aar3761}
  {10.1126/sciadv.aar3761} (\bibinfo {year} {2018})\BibitemShut {NoStop}%
\bibitem [{\citenamefont {Barreau}\ \emph {et~al.}(2020)\citenamefont
  {Barreau}, \citenamefont {Ross}, \citenamefont {Garg}, \citenamefont {Kraus},
  \citenamefont {Neumark},\ and\ \citenamefont {Leone}}]{Barreau2020}%
  \BibitemOpen
  \bibfield  {author} {\bibinfo {author} {\bibfnamefont {L.}~\bibnamefont
  {Barreau}}, \bibinfo {author} {\bibfnamefont {A.~D.}\ \bibnamefont {Ross}},
  \bibinfo {author} {\bibfnamefont {S.}~\bibnamefont {Garg}}, \bibinfo {author}
  {\bibfnamefont {P.~M.}\ \bibnamefont {Kraus}}, \bibinfo {author}
  {\bibfnamefont {D.~M.}\ \bibnamefont {Neumark}},\ and\ \bibinfo {author}
  {\bibfnamefont {S.~R.}\ \bibnamefont {Leone}},\ }\bibfield  {title} {\bibinfo
  {title} {{Efficient table-top dual-wavelength beamline for ultrafast
  transient absorption spectroscopy in the soft X-ray region}},\ }\bibfield
  {journal} {\bibinfo  {journal} {Scientific Reports}\ }\textbf {\bibinfo
  {volume} {10}},\ \href {https://doi.org/10.1038/s41598-020-62461-6}
  {10.1038/s41598-020-62461-6} (\bibinfo {year} {2020})\BibitemShut {NoStop}%
\bibitem [{\citenamefont {Jonas}\ \emph {et~al.}(2019)\citenamefont {Jonas},
  \citenamefont {Stiel}, \citenamefont {Gl{\"{o}}ggler}, \citenamefont {Dahm},
  \citenamefont {Dammer}, \citenamefont {Kanngie{\ss}er},\ and\ \citenamefont
  {Mantouvalou}}]{Jonas2019}%
  \BibitemOpen
  \bibfield  {author} {\bibinfo {author} {\bibfnamefont {A.}~\bibnamefont
  {Jonas}}, \bibinfo {author} {\bibfnamefont {H.}~\bibnamefont {Stiel}},
  \bibinfo {author} {\bibfnamefont {L.}~\bibnamefont {Gl{\"{o}}ggler}},
  \bibinfo {author} {\bibfnamefont {D.}~\bibnamefont {Dahm}}, \bibinfo {author}
  {\bibfnamefont {K.}~\bibnamefont {Dammer}}, \bibinfo {author} {\bibfnamefont
  {B.}~\bibnamefont {Kanngie{\ss}er}},\ and\ \bibinfo {author} {\bibfnamefont
  {I.}~\bibnamefont {Mantouvalou}},\ }\bibfield  {title} {\bibinfo {title}
  {{Towards Poisson noise limited optical pump soft X-ray probe NEXAFS
  spectroscopy using a laser-produced plasma source}},\ }\href
  {https://doi.org/10.1364/OE.27.036524} {\bibfield  {journal} {\bibinfo
  {journal} {Optics Express}\ }\textbf {\bibinfo {volume} {27}},\ \bibinfo
  {pages} {36524} (\bibinfo {year} {2019})}\BibitemShut {NoStop}%
\bibitem [{\citenamefont {Duris}\ \emph {et~al.}(2020)\citenamefont {Duris},
  \citenamefont {Li}, \citenamefont {Driver}, \citenamefont {Champenois},
  \citenamefont {MacArthur}, \citenamefont {Lutman}, \citenamefont {Zhang},
  \citenamefont {Rosenberger}, \citenamefont {Aldrich}, \citenamefont {Coffee},
  \citenamefont {Coslovich}, \citenamefont {Decker}, \citenamefont {Glownia},
  \citenamefont {Hartmann}, \citenamefont {Helml}, \citenamefont {Kamalov},
  \citenamefont {Knurr}, \citenamefont {Krzywinski}, \citenamefont {Lin},
  \citenamefont {Marangos}, \citenamefont {Nantel}, \citenamefont {Natan},
  \citenamefont {O'Neal}, \citenamefont {Shivaram}, \citenamefont {Walter},
  \citenamefont {Wang}, \citenamefont {Welch}, \citenamefont {Wolf},
  \citenamefont {Xu}, \citenamefont {Kling}, \citenamefont {Bucksbaum},
  \citenamefont {Zholents}, \citenamefont {Huang}, \citenamefont {Cryan},\ and\
  \citenamefont {Marinelli}}]{Duris2020}%
  \BibitemOpen
  \bibfield  {author} {\bibinfo {author} {\bibfnamefont {J.}~\bibnamefont
  {Duris}}, \bibinfo {author} {\bibfnamefont {S.}~\bibnamefont {Li}}, \bibinfo
  {author} {\bibfnamefont {T.}~\bibnamefont {Driver}}, \bibinfo {author}
  {\bibfnamefont {E.~G.}\ \bibnamefont {Champenois}}, \bibinfo {author}
  {\bibfnamefont {J.~P.}\ \bibnamefont {MacArthur}}, \bibinfo {author}
  {\bibfnamefont {A.~A.}\ \bibnamefont {Lutman}}, \bibinfo {author}
  {\bibfnamefont {Z.}~\bibnamefont {Zhang}}, \bibinfo {author} {\bibfnamefont
  {P.}~\bibnamefont {Rosenberger}}, \bibinfo {author} {\bibfnamefont {J.~W.}\
  \bibnamefont {Aldrich}}, \bibinfo {author} {\bibfnamefont {R.}~\bibnamefont
  {Coffee}}, \bibinfo {author} {\bibfnamefont {G.}~\bibnamefont {Coslovich}},
  \bibinfo {author} {\bibfnamefont {F.~J.}\ \bibnamefont {Decker}}, \bibinfo
  {author} {\bibfnamefont {J.~M.}\ \bibnamefont {Glownia}}, \bibinfo {author}
  {\bibfnamefont {G.}~\bibnamefont {Hartmann}}, \bibinfo {author}
  {\bibfnamefont {W.}~\bibnamefont {Helml}}, \bibinfo {author} {\bibfnamefont
  {A.}~\bibnamefont {Kamalov}}, \bibinfo {author} {\bibfnamefont
  {J.}~\bibnamefont {Knurr}}, \bibinfo {author} {\bibfnamefont
  {J.}~\bibnamefont {Krzywinski}}, \bibinfo {author} {\bibfnamefont {M.~F.}\
  \bibnamefont {Lin}}, \bibinfo {author} {\bibfnamefont {J.~P.}\ \bibnamefont
  {Marangos}}, \bibinfo {author} {\bibfnamefont {M.}~\bibnamefont {Nantel}},
  \bibinfo {author} {\bibfnamefont {A.}~\bibnamefont {Natan}}, \bibinfo
  {author} {\bibfnamefont {J.~T.}\ \bibnamefont {O'Neal}}, \bibinfo {author}
  {\bibfnamefont {N.}~\bibnamefont {Shivaram}}, \bibinfo {author}
  {\bibfnamefont {P.}~\bibnamefont {Walter}}, \bibinfo {author} {\bibfnamefont
  {A.~L.}\ \bibnamefont {Wang}}, \bibinfo {author} {\bibfnamefont {J.~J.}\
  \bibnamefont {Welch}}, \bibinfo {author} {\bibfnamefont {T.~J.}\ \bibnamefont
  {Wolf}}, \bibinfo {author} {\bibfnamefont {J.~Z.}\ \bibnamefont {Xu}},
  \bibinfo {author} {\bibfnamefont {M.~F.}\ \bibnamefont {Kling}}, \bibinfo
  {author} {\bibfnamefont {P.~H.}\ \bibnamefont {Bucksbaum}}, \bibinfo {author}
  {\bibfnamefont {A.}~\bibnamefont {Zholents}}, \bibinfo {author}
  {\bibfnamefont {Z.}~\bibnamefont {Huang}}, \bibinfo {author} {\bibfnamefont
  {J.~P.}\ \bibnamefont {Cryan}},\ and\ \bibinfo {author} {\bibfnamefont
  {A.}~\bibnamefont {Marinelli}},\ }\bibfield  {title} {\bibinfo {title}
  {{Tunable isolated attosecond X-ray pulses with gigawatt peak power from a
  free-electron laser}},\ }\href {https://doi.org/10.1038/s41566-019-0549-5}
  {\bibfield  {journal} {\bibinfo  {journal} {Nature Photonics}\ }\textbf
  {\bibinfo {volume} {14}},\ \bibinfo {pages} {30} (\bibinfo {year} {2020})},\
  \Eprint {https://arxiv.org/abs/1906.10649} {arXiv:1906.10649} \BibitemShut
  {NoStop}%
\bibitem [{\citenamefont {Driver}\ \emph {et~al.}(2020)\citenamefont {Driver},
  \citenamefont {Li}, \citenamefont {Champenois}, \citenamefont {Duris},
  \citenamefont {Ratner}, \citenamefont {Lane}, \citenamefont {Rosenberger},
  \citenamefont {Al-Haddad}, \citenamefont {Averbukh}, \citenamefont {Barnard},
  \citenamefont {Berrah}, \citenamefont {Bostedt}, \citenamefont {Bucksbaum},
  \citenamefont {Coffee}, \citenamefont {Dimauro}, \citenamefont {Fang},
  \citenamefont {Garratt}, \citenamefont {Gatton}, \citenamefont {Guo},
  \citenamefont {Hartmann}, \citenamefont {Haxton}, \citenamefont {Helml},
  \citenamefont {Huang}, \citenamefont {Laforge}, \citenamefont {Kamalov},
  \citenamefont {Kling}, \citenamefont {Knurr}, \citenamefont {Lin},
  \citenamefont {Drive}, \citenamefont {MacArthur}, \citenamefont {Marangos},
  \citenamefont {Nantel}, \citenamefont {Natan}, \citenamefont {Obaid},
  \citenamefont {O'Neal}, \citenamefont {Shivaram}, \citenamefont {Schori},
  \citenamefont {Walter}, \citenamefont {{Li Wang}}, \citenamefont {Wolf},
  \citenamefont {Marinelli},\ and\ \citenamefont {Cryan}}]{Driver2020}%
  \BibitemOpen
  \bibfield  {author} {\bibinfo {author} {\bibfnamefont {T.}~\bibnamefont
  {Driver}}, \bibinfo {author} {\bibfnamefont {S.}~\bibnamefont {Li}}, \bibinfo
  {author} {\bibfnamefont {E.~G.}\ \bibnamefont {Champenois}}, \bibinfo
  {author} {\bibfnamefont {J.}~\bibnamefont {Duris}}, \bibinfo {author}
  {\bibfnamefont {D.}~\bibnamefont {Ratner}}, \bibinfo {author} {\bibfnamefont
  {T.~J.}\ \bibnamefont {Lane}}, \bibinfo {author} {\bibfnamefont
  {P.}~\bibnamefont {Rosenberger}}, \bibinfo {author} {\bibfnamefont
  {A.}~\bibnamefont {Al-Haddad}}, \bibinfo {author} {\bibfnamefont
  {V.}~\bibnamefont {Averbukh}}, \bibinfo {author} {\bibfnamefont
  {T.}~\bibnamefont {Barnard}}, \bibinfo {author} {\bibfnamefont
  {N.}~\bibnamefont {Berrah}}, \bibinfo {author} {\bibfnamefont
  {C.}~\bibnamefont {Bostedt}}, \bibinfo {author} {\bibfnamefont {P.~H.}\
  \bibnamefont {Bucksbaum}}, \bibinfo {author} {\bibfnamefont {R.}~\bibnamefont
  {Coffee}}, \bibinfo {author} {\bibfnamefont {L.~F.}\ \bibnamefont {Dimauro}},
  \bibinfo {author} {\bibfnamefont {L.}~\bibnamefont {Fang}}, \bibinfo {author}
  {\bibfnamefont {D.}~\bibnamefont {Garratt}}, \bibinfo {author} {\bibfnamefont
  {A.}~\bibnamefont {Gatton}}, \bibinfo {author} {\bibfnamefont
  {Z.}~\bibnamefont {Guo}}, \bibinfo {author} {\bibfnamefont {G.}~\bibnamefont
  {Hartmann}}, \bibinfo {author} {\bibfnamefont {D.}~\bibnamefont {Haxton}},
  \bibinfo {author} {\bibfnamefont {W.}~\bibnamefont {Helml}}, \bibinfo
  {author} {\bibfnamefont {Z.}~\bibnamefont {Huang}}, \bibinfo {author}
  {\bibfnamefont {A.}~\bibnamefont {Laforge}}, \bibinfo {author} {\bibfnamefont
  {A.}~\bibnamefont {Kamalov}}, \bibinfo {author} {\bibfnamefont {M.~F.}\
  \bibnamefont {Kling}}, \bibinfo {author} {\bibfnamefont {J.}~\bibnamefont
  {Knurr}}, \bibinfo {author} {\bibfnamefont {M.~F.}\ \bibnamefont {Lin}},
  \bibinfo {author} {\bibfnamefont {T.}~\bibnamefont {Drive}}, \bibinfo
  {author} {\bibfnamefont {J.~P.}\ \bibnamefont {MacArthur}}, \bibinfo {author}
  {\bibfnamefont {J.~P.}\ \bibnamefont {Marangos}}, \bibinfo {author}
  {\bibfnamefont {M.}~\bibnamefont {Nantel}}, \bibinfo {author} {\bibfnamefont
  {A.}~\bibnamefont {Natan}}, \bibinfo {author} {\bibfnamefont
  {R.}~\bibnamefont {Obaid}}, \bibinfo {author} {\bibfnamefont {J.~T.}\
  \bibnamefont {O'Neal}}, \bibinfo {author} {\bibfnamefont {N.~H.}\
  \bibnamefont {Shivaram}}, \bibinfo {author} {\bibfnamefont {A.}~\bibnamefont
  {Schori}}, \bibinfo {author} {\bibfnamefont {P.}~\bibnamefont {Walter}},
  \bibinfo {author} {\bibfnamefont {A.}~\bibnamefont {{Li Wang}}}, \bibinfo
  {author} {\bibfnamefont {T.~J.}\ \bibnamefont {Wolf}}, \bibinfo {author}
  {\bibfnamefont {A.}~\bibnamefont {Marinelli}},\ and\ \bibinfo {author}
  {\bibfnamefont {J.~P.}\ \bibnamefont {Cryan}},\ }\bibfield  {title} {\bibinfo
  {title} {{Attosecond transient absorption spooktroscopy: A ghost imaging
  approach to ultrafast absorption spectroscopy}},\ }\href
  {https://doi.org/10.1039/c9cp03951a} {\bibfield  {journal} {\bibinfo
  {journal} {Physical Chemistry Chemical Physics}\ }\textbf {\bibinfo {volume}
  {22}},\ \bibinfo {pages} {2704} (\bibinfo {year} {2020})},\ \Eprint
  {https://arxiv.org/abs/1909.07441} {arXiv:1909.07441} \BibitemShut {NoStop}%
\end{thebibliography}%

\end{document}